\documentclass[11pt,a4paper]{article}
\usepackage{jheppub}
\begin{document} 
%\end{document}
%\topmargin 0.30cm 
%\addtolength{\topmargin}{-0.5cm} 
%%%

%\relax
%%%%%%%%%%%
\newcommand{\tr}{\mbox{tr}}
\def\beq{\begin{equation}}
\def\eeq{\end{equation}}
\def\bea{\begin{array}}
\def\eea{\end{array}}
\def\be{\begin{equation}}
\def\ee{\end{equation}}
\def\ba{\begin{eqnarray}}
\def\ea{\end{eqnarray}}
\def\pbarp{ \bar{{\rm p}} {\rm p} }
\def\pp{ {\rm p} {\rm p} }
\def\ifb{ {\rm fb}^{-1} }
\def\del{\partial }
\def\to{\rightarrow}
\def\To{\Rightarrow}
\def\tolr{\leftrightarrow}
\def\dis{\displaystyle}
\def\f{\frac}
\def\[{\left[}
\def\]{\right]}
\def\({\left(}
\def\){\right)}
\def\LL{{\left\lgroup}}
\def\RR{{\right\rgroup}}
\def\nn{{/\hspace*{-1.83mm}}}
%%%%%
%%%%%%%%%%%%%%%%%%%%%%%%%%%%%%%%%%%%%%%%%%%%%%%%%%%%%%%%
\def\la{{\lambda}}
\def\ep{{\epsilon}}
\def\qs{{\sqrt{2}}}
\def\cxx{{c_x}}
\def\sxx{{s_x}}
\def\cyy{{c_y}}
\def\syy{{s_y}}
\def\Ah{{\widehat{A}}}
\def\ms{{\widetilde{m}}}
\def\sm0{{\widetilde{m}_0}}
\def\sB{{\sin\beta}}
\def\cB{{\cos\beta}}
\def\tanB{{\tan\beta}}
\def\cotB{{\cot\beta}}
\def\sM{{\widetilde{\cal M}}}
\def\sQ{{\widetilde{Q}}}
\def\sU{{\widetilde{U}}}
\def\sD{{\widetilde{D}}}
\def\sL{{\widetilde{L}}}
\def\sE{{\widetilde{E}}}
\def\Hbc{H^\pm bc}
\def\bs{\tilde{b}}
\def\gs{\tilde{g}}
\def\ts{\tilde{t}}
\def\cs{\tilde{c}}
\def\us{\tilde{u}}
\def\cut{{\Lambda}}
\def\HH{{\widetilde{H}}}
\def\K{\kappa}
\def\ov{\overline}
\def\LL{{\cal L}}
\def\U1em{{U(1)_{\rm em}}}
\def\to{\rightarrow}
\def\To{\Rightarrow}
\def\G{{\cal G}}
\def\GSM{{\cal G}_{\rm SM}}
\def\CL{{\cal C}_L}
\def\CR{{\cal C}_R}
\def\sb{s_{\beta}}
\def\sq2{\sqrt{2}}
\def\gaga{\gamma\gamma}
\def\tanb{\tan\hspace*{-1mm}\beta}
\def\ee{e^+e^-}
\def\MM{M_{\star}}
\def\pbinv{pb$^{-1}$}
\def\fbinv{fb$^{-1}$}
\def\lumi{\int\!{\cal L}}
\def\STU{{(S,T,U)}}
%\def\Zbb{{Zb\bar{b}}}
%\def\ca{c_{\alpha}}
%\def\sa{s_{\alpha}}
%\def\End{\end{document}}
%%%
%%% math-def:
\newcommand{\lae}{\stackrel{<}{\sim}}
\newcommand{\gae}{\stackrel{>}{\sim}}
\newcommand{\gsim}{\mbox{ \raisebox{-1.0ex}{$\stackrel{\textstyle >}
{\textstyle \sim}$ }}}
\newcommand{\lsim}{\mbox{ \raisebox{-1.0ex}{$\stackrel{\textstyle <}
{\textstyle \sim}$ }}}
%ccccccc csk
\def\Journal#1#2#3#4{{#1} {\bf #2} (#4) #3}
% Some useful journal names
\def\NCA{\em Nuovo Cimento}
\def\NIM{\em Nucl. Instrum. Methods}
\def\NIMA{{\em Nucl. Instrum. Methods} A}
\def\NPB{{\em Nucl. Phys.} B}
\def\PLB{{\em Phys. Lett.}  B}
\def\PRL{\em Phys. Rev. Lett.}
\def\PRD{{\em Phys. Rev.} D}
\def\ZPC{{\em Z. Phys.} C}
\def\EPC{{\em Euro. Phys. J.} C}
\def\PTP{\em Prog.~Theor.~Phys.}
\def\ibid{\it ibid.}
%ccccc
%
%\def\fsl#1{\setbox0=\hbox{$#1$}                 % set a box for #1 
%   \dimen0=\wd0                                 % and get its size
%   \setbox1=\hbox{/} \dimen1=\wd1               % get size of /
%   \ifdim\dimen0>\dimen1                        % #1 is bigger
%      \rlap{\hbox to \dimen0{\hfil/\hfil}}      % so center / in box
%      #1                                        % and print #1
%   \else                                        % / is bigger
%      \rlap{\hbox to \dimen1{\hfil$#1$\hfil}}   % so center #1
%      /                                         % and print /
%   \fi}  
%\newcommand{\VEV}[1]{\langle #1 \rangle}
%
%%
%
%\def\thisday{June, 2008} %~and~ hep-ph/yymmnnn~~} 
%
%%%%%%%%%%%%%%%%%%%%%%%%%%%%%%%%%%%%%%%%%%%%%%%%%
%\draft
%@twocolumnfalse\endcsname
\title{Decoupling property of the supersymmetric Higgs sector 
with four doublets}%
%\preprint{KANAZAWA-11-13, UT-HET 056, KU-PH-009}
%  
\author[a]{Mayumi Aoki, }
\author[b]{ Shinya Kanemura,}
\author[c]{Tetsuo Shindou,}
\author[b]{and  Kei Yagyu}

\affiliation[a]{Institute~for~Theoretical~Physics,~Kanazawa~University,\\
Kanazawa~920-1192,~Japan}
\affiliation[b]{Department of Physics, University of Toyama, \\
3190 Gofuku, Toyama 930-8555, Japan}
\affiliation[c]{Division of Liberal Arts, Kogakuin University, \\
1-24-2 Shinjuku, Tokyo 163-8677, Japan}

\emailAdd{mayumi@hep.s.kanazawa-u.ac.jp}
\emailAdd{kanemu@sci.u-toyama.ac.jp}
\emailAdd{shindou@cc.kogakuin.ac.jp}
\emailAdd{keiyagyu@jodo.sci.u-toyama.ac.jp}
\abstract{
 In supersymmetric standard models with multi Higgs doublet fields, 
 selfcoupling constants in the Higgs potential come only from the D-terms
 at the tree level.
 We investigate the decoupling property of additional two heavier Higgs
 doublet fields in the supersymmetric standard model with four Higgs doublets.
 In particular, we study how they can modify the predictions
 on the quantities well predicted in the minimal supersymmetric standard
 model (MSSM), when the extra doublet fields are rather heavy to be 
 measured at collider experiments.  
 The B-term mixing between these extra heavy Higgs bosons
 and the relatively light MSSM-like Higgs bosons can significantly change
 the predictions in the MSSM such as on the masses of MSSM-like
 Higgs bosons as well as the mixing angle for the two light CP-even scalar states. 
 We first give formulae for deviations in the observables of the MSSM   
 in the decoupling region for the extra two doublet fields.  
 We then examine possible deviations in the Higgs sector 
 numerically, and discuss their phenomenological implications.}
 
%\pacs{\, 14.80.Da, 12.60.Fr, 12.60.Jv }%\hfill~~[\today] }
%\arxivnumber{1108.1356}
%\notoc
%\toccontinuoustrue
%\begin{document}    
%\begin{document}    
      
\maketitle

\section{Introduction}

Although the standard model (SM) has enjoyed a marvelous success 
in explaining phenomena observed at collider experiments, the  
physics of electroweak symmetry breaking remains unknown. 
The experimental detection of the Higgs boson is the most
important issue to confirm the standard picture for the origin
of mass of particles, and the Higgs boson search is underway
at the Fermilab Tevatron and the CERN LHC.
On the other hand, the Higgs sector of the SM is known to 
cause the hierarchy problem~\cite{hierarchy} due to the quadratic 
divergence in the one-loop correction to the mass of the Higgs boson.
In addition, there are several phenomena
confirmed by the experiments which cannot 
be understood within the SM,
such as the neutrino oscillation, the existence of 
dark matter and the baryon asymmetry of the Universe. 
Therefore, the SM must be extended to solve these problems.

Supersymmetry (SUSY) is expected to be a good candidate of 
new physics.  It can solve the hierarchy problem by the consequence of
the nonrenormalization theorem~\cite{nonreno-theorem}.
The stabilized Higgs boson mass makes it possible to directly connect
the electroweak scale with very high scales such as the Planck scale or 
that of grand unification.
Supersymmetric extensions of the SM
with the R parity also provide dark matter candidates~\cite{SUSY-DM}. 
In addition, various mechanisms of generating tiny neutrino
masses~\cite{SS-I,SS-II, SS-III, SS-R,SS-R2} 
as well as those of baryogenesis~\cite{B-AD,B-Lepto,B-EW}
may also be compatible to supersymmetric models. 

The minimal supersymmetric standard model (MSSM) is 
a SUSY extension of the SM with the minimal number of particle content. 
It requires two Higgs doublet fields for cancellation of anomaly. 
The most striking phenomenological prediction of the model 
is that on the mass ($m_h$) of the lightest CP-even Higgs boson $h$.
It can be calculated to be less than the mass of the Z boson 
at the tree level. Such an upper bound on $m_h$   
comes from the fact that the interaction terms in the 
Higgs potential are given only by D-term contributions
which are determined by the gauge coupling constants. 
At the one-loop level the trilinear top-Yukawa term in the
superpotential gives a significant F-term contribution to
$m_h$~\cite{mh-MSSM,mh-MSSM1,dabelstein}, by which $m_h$ can be above
the lower bound from the direct search results at the CERN LEP
experiment~\cite{h-search-LEP}.
The calculation has been improved with higher order
corrections~\cite{mh-MSSM2-rge,mh-MSSM2,mh-MSSM3}.
Apart from $m_h$, the masses of $H$, $H^\pm$ and
the mixing angle $\alpha$ are a function of only two
input parameters at the tree level; i.e., $m_A$ and $\tan\beta$,
where $m_A$ is the mass of CP-odd Higgs boson $A$, $H$ is the heavier
CP-even Higgs boson, $H^\pm$ are the charged Higgs bosons,
$\tan\beta$ is the ratio of vacuum expectation values (VEVs) of
the two Higgs bosons and
$\alpha$ is the angle which diagonalizes the CP-even scalar states.
In particular, there is a simple tree-level relationship among the masses of 
$H^\pm$, $A$ and the W boson $W^\pm$ as $m_{H^\pm}^2=m_A^2+m_W^2$,
where $m_{H^\pm}$ and $m_W$ are respectively the masses of $H^\pm$ and $W^\pm$.
These characteristic predictions can be used
to confirm the MSSM. 

If by experiments the Higgs boson is found and its mass turns out 
to be higher than the upper bound predicted in the MSSM,
we must abandon the model.
It however does not necessarily imply that the SUSY itself is excluded.
For example, in the next-to-minimal supersymmetric standard model
(NMSSM), where a gauge singlet chiral superfield $S$ is added to the MSSM, 
the upper bound on $m_h$ is greater than that in the MSSM
due to  the F-term contribution from the trilinear
$\lambda H_1 H_2 S$ interaction in the superpotential~\cite{mh-NMSSM,mh-NMSSM2},
where $H_1$ and $H_2$ are two chiral superfields for the Higgs doublets. 
The effect of this term can gain the mass upper bound, so that  
$m_h$ can be as large as $140$ GeV ($400$ GeV)  
by assuming that the running coupling constant $\lambda$ does not
blow up below the grand unification scale $\sim 10^{16}$ GeV 
(the TeV scale) around $\tan\beta \sim 2$~\cite{mh-NMSSM2}. 
The similar effect on $m_h$ also appears
in the model with additional triplet chiral superfields
with the hypercharge of $Y=0$ or $Y=\pm 1$~\cite{mh-triplet,ksy}.
The other possibility to change the MSSM prediction on 
$m_h$ is to introduce extra gauge symmetries
which are broken spontaneously at the TeV scale.  
The mixing of gauge bosons of these new symmetries 
with those of the SM gauge symmetries yields
the new D-term contribution to the Higgs potential, 
which can change for instance $m_h$~\cite{barbieri,add-gauge}. 

In general, interaction terms of the Higgs potential in extended SUSY standard models 
are composed of the D-term and the F-term as well as the trilinear
soft-breaking term, so that extended SUSY Higgs sectors predict
different masses and coupling constants from those in the MSSM. 
If the extra scalar fields are sufficiently light, they largely
mix with the MSSM-like Higgs bosons,
and consequently predictions on the MSSM observables are modified from
the MSSM values at the tree level.
In addition, such light extra scalars may directly be detected at the
experiments. In this case the model can be easily distinguished
from the MSSM. 
On the other hand,  extra scalar bosons
may be too heavy to be measured directly. 
Even in such a case the effect of these extra fields can change
predictions on masses and coupling constants
for the MSSM-like scalar bosons significantly at the low energy. 
This is not contradict with the decoupling theorem by Appelquist 
and Carazzone~\cite{dec-theorem}. In fact, in the large mass limit of all the 
fields other than those in the SM, the model can be continuously 
reduced to the SM. 

In this paper, 
we consider the model (4HDSSM) where two extra doublet superfields 
are added into the MSSM in order to examine the decoupling 
property of extended Higgs sectors without interactions
from the F-term at the tree level. 
In this model,  
if there are mixings between the MSSM-like doublet fields and
the extra doublet fields due to a large soft-breaking B-term mixing,
nonvanishing effects can appear in the MSSM observables such as
$m_h$, $m_H$, $m_{H^\pm}$ and $\alpha$.
Notice that this nonvanishing effect is not the so-called nondecoupling effect.
The usual nondecoupling effect, which for example appears
in the radiative corrections to the gauge boson two-point functions, 
does not vanish in the SM-like limit with taking the extra particles to
be heavy.
On the other hand, the present effect due to the large soft-breaking
B-term mixing on MSSM observables in the 4HDSSM
can be significant in the MSSM-like limit but does decouple in the
SM-like limit by taking $m_A \to \infty$ according to the
decoupling theorem~\cite{dec-theorem}. 
Therefore, we call such a nonvanishing effect on the MSSM observables
as the quasi-nondecoupling effect. 
We deduce analytic expressions of such quasi-nondecoupling effects 
on the MSSM observables when the extra doublet fields are heavy.  
Modifications from the MSSM predictions by these effects
are then studied numerically. 
We find that the quasi-nondecoupling effect due to the B-term mixing can be
significant.
For example, $m_h$ can be larger than about $10$ \% as
compared to that of the MSSM prediction with the same MSSM input parameters,
while $m_{H^\pm}$ can be about $20$ \% ($10$ \%) smaller than the MSSM
prediction when $m_A=150$ GeV ($200$ GeV).
The mixing angle $\sin(\beta-\alpha)$ also receives
a significant deviation from the MSSM value.
The Higgs potential and the mass matrices in the 4HDSSM 
have been analyzed by Gupta and Wells~\cite{Gupta}, but they 
have not included these quasi-nondecoupling effects discussed here.
    
Suppose that the candidates of SUSY partner particles and 
extra Higgs bosons in the MSSM ($H$, $A$ and $H^\pm$)
are found  at the LHC or the
International Linear Collider (ILC) in future  
and that their properties look like consistent with the MSSM. 
In such a case, to confirm whether it is really of the MSSM 
or not, the MSSM relations are tested as accurately as possible 
by experiments. 
Our study can be particularly important in this case. 

 In general, in extended Higgs sectors with multi-doublets, 
there necessarily appears the flavor changing neutral current
(FCNC),  which is strictly constrained by experiments. 
A simple prescription to avoid it would be
imposing the (softly-broken) discrete $Z_2$ symmetry to the
model~\cite{fcnc}. We classify the type of Yukawa interaction under the discrete
symmetry according to the assignment of the $Z_2$ charges.
Phenomenology of the 4HDSSM has been discussed by Marshall and Sher
in the case with the lepton specific Yukawa interaction~\cite{sher}.
We here  do not discuss influence of the heavy extra two doublet fields
on the flavor physics, just assuming that the tree-level FCNC is forbidden 
by a $Z_2$ symmetry. 
We then concentrate on the deviations in the Higgs sector 
due to the quasi-nondecoupling effects of these extra doublets. 

This paper is organized as follows. 
In Sec.~II, the 4HDSSM is defined
and the mass matrices for the Higgs bosons are obtained. 
In Sec.~III, we give a short description of the decoupling property in
extended SUSY Higgs sectors, where the (quasi-)nondecoupling effect
is discussed. We then present formulae for the
quasi-nondecoupling effects on the MSSM observables.
Numerical studies are also shown, and the results are discussed. 
Conclusions are given in Sec.~IV.  
We show the method of obtaining the Higgs potential in the useful basis
from the general basis in Appendix~A. 
The flavor structure in the 4HDSSM is shortly discussed in Appendix~B.

\section{The model}

%\subsection{Lagrangian}

We here discuss the 4HDSSM, in which two extra isospin-doublet chiral superfields 
$H_3$ ($Y=-1/2$) and $H_4$ ($Y=1/2$) are introduced to 
the MSSM in addition to the Higgs doublets $H_1$ ($Y=-1/2$) and $H_2$ ($Y=1/2$). 
The general expression for the superpotential with the R parity
is given in terms of chiral superfields as 
\begin{align}
W=&
-(\hat{Y}_u)_{ij}U_{Ri}^cH_2\cdot Q_{Lj}
+(\hat{Y}_d)_{ij}D_{Ri}^cH_1\cdot Q_{Lj}
+(\hat{Y}_e)_{ij}E_{Ri}^cH_1\cdot L_{Lj}
\nonumber\\
&-(\hat{Y}_u^{\prime})_{ij}U_{Ri}^cH_4\cdot Q_{Lj}
+(\hat{Y}_d^{\prime})_{ij}D_{Ri}^cH_3\cdot Q_{Lj}
+(\hat{Y}_e^{\prime})_{ij}E_{Ri}^cH_3\cdot L_{Lj}
\nonumber\\
&-\mu_{12}H_1\cdot H_2
-\mu_{14}H_1\cdot H_4
-\mu_{32}H_3\cdot H_2
-\mu_{34}H_3\cdot H_4\;, \label{eq:superpotential}
\end{align}
where $Q_{Li}$, $U_{Ri}^c$ ($D_{Ri}^c$) are chiral superfields for the $i$-th
generation left-handed quark doublet and right-handed up-type (down-type)
quark singlet while $L_{Li}$ and $E_{Ri}^c$ are those for the $i$-th generation   
left-handed lepton doublet and right-handed charged lepton singlet,
respectively.  
The most general holomorphic soft-SUSY-breaking terms with the R parity is 
\begin{align}
\mathcal{L}_{\mbox{soft}}=&
-\frac{1}{2}(M_1^{}\tilde{B}\tilde{B}
+M_2^{}\tilde{W}\tilde{W}
+M_3^{}\tilde{G}\tilde{G})
\displaybreak[0]
\nonumber\\
&-
\left((\tilde{M}_{\tilde{q}}^2)_{ij}\tilde{q}_{Li}^{\dagger}\tilde{q}_{Lj}^{}
+(\tilde{M}_{\tilde{u}}^2)_{ij}\tilde{u}_{Ri}^*\tilde{u}_{Rj}^{}
+(\tilde{M}_{\tilde{d}}^2)_{ij}\tilde{d}_{Ri}^*\tilde{d}_{Rj}^{}
+(\tilde{M}_{\tilde{\ell}}^2)_{ij}\tilde{\ell}_{Li}^{\dagger}\tilde{\ell}_{Lj}^{}
+(\tilde{M}_{\tilde{e}}^2)_{ij}\tilde{e}_{Ri}^*\tilde{e}_{Rj}^{}\right)
\displaybreak[0]
\nonumber\\
&-\left(
(\tilde{M}_{-}^2)_{11}^{}\hat{\Phi}_1^{\dagger}\hat{\Phi}_1^{}
+(\tilde{M}_{-}^2)_{13}^{}\hat{\Phi}_1^{\dagger}\hat{\Phi}_3^{}
+(\tilde{M}_{-}^2)_{13}^*\hat{\Phi}_3^{\dagger}\hat{\Phi}_1^{}
+(\tilde{M}_{-}^2)_{33}^{}\hat{\Phi}_3^{\dagger}\hat{\Phi}_3^{}\right.\nonumber\\
&\left.\phantom{S}
+(\tilde{M}_{+}^2)_{22}^{}\hat{\Phi}_2^{\dagger}\hat{\Phi}_2^{}
+(\tilde{M}_{+}^2)_{24}^{}\hat{\Phi}_2^{\dagger}\hat{\Phi}_4^{}
+(\tilde{M}_{+}^2)_{24}^{*}\hat{\Phi}_4^{\dagger}\hat{\Phi}_2^{}
+(\tilde{M}_{+}^2)_{44}^{}\hat{\Phi}_4^{\dagger}\hat{\Phi}_4^{}
\right)
\displaybreak[0]\nonumber\\
&-\left(
-(A_{u})^{ij}\tilde{u}_{Ri}^* \hat{\Phi}_2\cdot \tilde{q}_{Lj}
-(A_{u}^{\prime})^{ij}\tilde{u}_{Ri}^* \hat{\Phi}_4\cdot \tilde{q}_{Lj}
+(A_{d})^{ij}\tilde{d}_{Ri}^* \hat{\Phi}_1\cdot \tilde{q}_{Lj}
+(A_{d}^{\prime})^{ij}\tilde{d}_{Ri}^* \hat{\Phi}_3\cdot \tilde{q}_{Lj}\right.
\displaybreak[0]\nonumber\\
&\left.\phantom{Space}
+(A_{e})^{ij}\tilde{e}_{Ri}^* \hat{\Phi}_1\cdot \tilde{\ell}_{Lj}
+(A_{e}^{\prime})^{ij}\tilde{e}_{Ri}^* \hat{\Phi}_3\cdot \tilde{\ell}_{Lj}
+h.c.\right)
\displaybreak[0]\nonumber \\
&-\left(B_{12}\mu_{12} \hat{\Phi}_1^{}\cdot \hat{\Phi}_2^{}
+B_{34}\mu_{34} \hat{\Phi}_3^{}\cdot \hat{\Phi}_4^{}
+B_{14}\mu_{14} \hat{\Phi}_1^{}\cdot \hat{\Phi}_4^{}
+B_{32}\mu_{32} \hat{\Phi}_3^{}\cdot \hat{\Phi}_2^{}
+\text{h.c.}\right)\;, \label{soft-term}
\end{align}
where $\tilde{B}$, $\tilde{W}$ and $\tilde{G}$ are gauginos
corresponding to the SM gauge symmetries of $U(1)$, $SU(2)$ and $SU(3)$,
respectively, $\tilde{q}_{Li}$, $\tilde{u}_{Ri}^\ast$, $\tilde{d}_{Ri}^\ast$,
$\tilde{\ell}_{Li}$ and $\tilde{e}_{Ri}^\ast$  are respectively
the scalar component fields of
$Q_{Li}$, $U_{Ri}^c$, $D_{Ri}^c$, $L_{Li}$ and $E_{Ri}^c$, and
$\hat{\Phi}_j$ ($j=1$-$4$) are the scalar component doublet fields
of the chiral superfields $H_j$.
From $W$ and $\mathcal{L}_{\text{soft}}^{}$, the Lagrangian is constructed as 
\begin{align}
\mathcal{L}=&
\mathcal{L}_{\text{kinetic}}
+\mathcal{L}_{\text{gauge--matter}}
-\left(\frac{1}{2}\frac{\partial^2 W}{\partial \varphi_i\partial\varphi_j}
\psi_{Li}\cdot\psi_{Lj}+h.c.\right)
-\frac{1}{2}(g_a)^2
(\varphi^*_{i}T^a_{ij}\varphi_{j})^2
- \left|\frac{\partial W}{\partial \varphi_i}\right|^2 \nonumber\\ 
&+\mathcal{L}_{\mbox{soft}}
\;, \label{eq:lag}
\end{align}
where $\psi_{Li}$ and $\varphi_j$ represent
fermion and scalar component fields of chiral superfields in the model.

\begin{table}[t]
\begin{center}
{\renewcommand\arraystretch{1}
\begin{tabular}{|c||c|c|c|c|c|c|c|c|c||c|}\hline
      &$H_1$&$H_2$&$H_3$&$H_4$&$U_R^c$&$D_R^c$&$E_R^c$&$Q_L$&$L_L$& $N_R^c$ \\\hline\hline
Type A&$+$  &$+$  &$-$  &$-$  &$+$&$+$&$+$&$+$&$+$& $+$   \\\hline
Type B&$+$  &$+$  &$-$  &$-$  &$+$&$+$&$-$&$+$&$+$& $+$   \\\hline
Type C&$+$  &$+$  &$-$  &$-$  &$+$&$+$&$+$&$+$&$+$& $-$   \\\hline
Type D&$+$  &$+$  &$-$  &$-$  &$+$&$+$&$-$&$+$&$+$& $-$   \\\hline
\end{tabular}}
\caption{Classification for the charge assignment for
 the $Z_2$ symmetry in the 4HDSSM. Type C and Type D are introduced only when $N_{Ri}^c$
 are added to the model.}
\label{z2}
\end{center}
\end{table}

There are two Higgs doublets for each quantum number, so that they can mix with each other. 
The Yukawa sector then produces a dangerous FCNC via the scalar boson
exchange at the tree level.
There are several ways to eliminate such an excessive FCNC.
In non-SUSY extended Higgs sectors with multi-doublets,
a softly-broken discrete $Z_2$ symmetry is often imposed~\cite{fcnc}. 
In the general two Higgs doublet model, there are four 
types of Yukawa interactions under such a $Z_2$ symmetry 
depending on the assignment of the $Z_2$ charge~\cite{barger,grossman,typeX}.
The other possibility of eliminating the FCNC may be to
consider a certain of alignment in the Yukawa sector~\cite{pich}, but
we do not consider this possibility  in this paper. 
In the 4HDSSM, we also impose the $Z_2$ symmetry to eliminate the FCNC. 
There are two types of Yukawa interactions (Type A and Type B)
as shown in Table~1, assuming that all the Higgs doublet fields receive VEVs.
If we introduce additional chiral superfields $N_{Ri}^c$ for
right-handed neutrinos which are singlet under the SM gauge symmetries,
possible number of the type of Yukawa interaction becomes doubled
under the $Z_2$ symmetry, depending on the two possible 
assignment of the $Z_2$ charge for $N_{Ri}^c$.
We define additional two types in Table 1 (Type C and Type D) which
correspond to the $Z_2$ odd $N_{Ri}^c$.
%We summarize the types of Yukawa interaction in the 4HDSSM in Table 1.
%
Under the $Z_2$ symmetry,
some of the Yukawa coupling constants are forbidden for each type of Yukawa interaction.
For example, in the MSSM-like Yukawa interaction (Type A)
$\hat{Y}_u'=\hat{Y}_d'=\hat{Y}_e'=0$ is required,
while in the lepton specific one (Type B) we have $\hat{Y}_u'=\hat{Y}_d'=\hat{Y}_e=0$.
Marshall and Sher discussed phenomenology of the Type B Yukawa interaction
in the 4HDSSM~\cite{sher}. 
Notice that the dimensionful parameters are not forbidden as long
as the discrete symmetry is softly broken~\cite{comment}.
In this paper, we assume that the FCNC is sufficiently suppressed
by a softly-broken $Z_2$ symmetry. 
However, we do not specify the type of Yukawa interaction, because 
all the essential results in this paper 
do not depend on the types of Yukawa interaction.
The general discussion on phenomenological consequences 
in flavor physics is given elsewhere~\cite{aksy-flavor}.

%\subsection{The Higgs potential}

From the Lagrangian in Eq.~(\ref{eq:lag}), we can extract the Higgs potential of the model,
in which each neutral scalar component $\hat{\Phi}_i$ ($i=1$-$4$) receives
the VEV. However, because $H_1$ and $H_3$ ($H_2$ and $H_4$)
have the same quantum numbers under the $SU(2)\times U(1)$ gauge
symmetries, there are $U(2)$ symmetries in the D-terms $[(g_a)^2
(\varphi_\alpha^\ast T^a_{\alpha\beta} \varphi_\beta)^2]$ in the
potential. By using the $U(2)$ symmetry, we may rotate the
fields $\hat{\Phi}_1$ and $\hat{\Phi}_3$ and take the
basis in which only one of the doublets receives the VEV while
the other does not.
We also can perform the same procedure for $\hat{\Phi}_2$ and
$\hat{\Phi}_4$: see Appendix A for details. 
Consequently, without loss of generality we can rewrite the
Higgs potential as 
\begin{align}
V_{\text{H}}^{}=&
\begin{pmatrix}
\Phi_1^{\dagger}&\Phi_1^{\prime\dagger}
\end{pmatrix}
\begin{pmatrix}
(M_{1}^{2})_{11}^{}&
(M_{1}^{2})_{12}^{}\\
(M_{1}^{2})_{12}^{*}&
(M_{1}^{2})_{22}^{}\\
\end{pmatrix}
\begin{pmatrix}
\Phi_1^{}\\ \Phi_1^{\prime}
\end{pmatrix}
%\displaybreak[0]
%\nonumber\\
%&
+
\begin{pmatrix}
\Phi_2^{\dagger}&\Phi_2^{\prime\dagger}
\end{pmatrix}
\begin{pmatrix}
(M_{2}^{2})_{11}^{}&
(M_{2}^{2})_{12}^{}\\
(M_{2}^{2})_{12}^{*}&
(M_{2}^{2})_{22}^{}\\
\end{pmatrix}
\begin{pmatrix}
\Phi_2^{} \\ \Phi_2^{\prime}
\end{pmatrix}
\displaybreak[0]
\nonumber\\
&
-\left(
\begin{pmatrix}
{\Phi}_1^{}&
{\Phi}_1^{\prime}
\end{pmatrix}
\begin{pmatrix}
(M_{3}^2)_{11}&
(M_{3}^2)_{12}\\
(M_{3}^2)_{21}&
(M_{3}^2)_{22}&
\end{pmatrix}
\cdot 
\begin{pmatrix}
{\Phi}_2^{}\\
{\Phi}_2^{\prime}
\end{pmatrix}
+\text{h.c.}\right)
\displaybreak[0]
\nonumber\\
&
+\frac{g^{\prime 2}+g^2}{8}\left(
\Phi_2^{\dagger}\Phi_2^{}
+\Phi_2^{\prime\dagger}\Phi_2^{\prime}
-\Phi_1^{\dagger}\Phi_1^{}
-\Phi_1^{\prime\dagger}\Phi_1^{\prime}
\right)^2
\displaybreak[0]
\nonumber\\
&
+\frac{g^2}{2}\left\{
(\Phi_1^{\dagger}\Phi_2^{})(\Phi_2^\dagger\Phi_1^{})
+(\Phi_1^{\dagger}\Phi_2^{\prime})(\Phi_2^{\prime\dagger}\Phi_1^{})
+(\Phi_1^{\prime\dagger}\Phi_2^{})(\Phi_2^\dagger\Phi_1^{\prime})
+(\Phi_1^{\prime\dagger}\Phi_2^{\prime})(\Phi_2^{\prime\dagger}\Phi_1^{\prime})
\right.
\nonumber\\
&\phantom{+\frac{g^2}{2}()()}\left.
+(\Phi_1^{\dagger}\Phi_1^{\prime})(\Phi_1^{\prime\dagger}\Phi_1^{})
-(\Phi_1^{\dagger}\Phi_1^{})(\Phi_1^{\prime\dagger}\Phi_1^{\prime})
+(\Phi_2^{\dagger}\Phi_2^{\prime})(\Phi_2^{\prime\dagger}\Phi_2^{})
-(\Phi_2^{\dagger}\Phi_2^{})(\Phi_2^{\prime\dagger}\Phi_2^{\prime})
\right\}\;,  \label{V_gi}
\end{align}
where $\Phi_1$ ($Y=-1/2$) and $\Phi_2$ ($Y=1/2$) have VEVs, while
$\Phi_1'$ ($Y=-1/2$) and $\Phi_2'$ ($Y=1/2$) do not.
Throughout this paper, we restrict ourselves in the CP invariant case. 
We thus  hereafter neglect all CP violating phases in the dimentionful parameters. 

  %\subsection{Vacuum conditions}

The rotated Higgs doublet fields $\Phi_1$,
$\Phi_1'$, $\Phi_2$ and $\Phi_2'$
are expressed as 
\begin{align}
 \Phi_1=\left[
 \begin{array}{c}
    \varphi_1^{0\ast} \\
    -\varphi_1^-\\
  \end{array}
 \right],
\quad 
 \Phi_2=\left[
 \begin{array}{c}
    \varphi_2^+ \\
    \varphi_2^0\\
  \end{array}
 \right],
\quad
 \Phi_1^\prime=\left[
 \begin{array}{c}
    \varphi_1^{\prime 0\ast} \\
    -\varphi_1^{\prime -}\\
  \end{array}
 \right],
\quad
 \Phi_2^\prime=\left[
 \begin{array}{c}
    \varphi_2^{\prime +} \\
    \varphi_2^{\prime 0}\\
  \end{array}
 \right],
\end{align}
where the neutral scalar fields can be parameterized as 
\begin{align}
&\varphi_1^0 = \frac{1}{\sqrt{2}}\left(v_1^{}+\phi_1^{}+i\chi_1^{}\right)\;,\quad
\varphi_2^0 = \frac{1}{\sqrt{2}}\left(v_2^{}+\phi_2^{}+i\chi_2^{}\right)\;,\quad
\nonumber\\
&\varphi_1^{\prime 0} = \frac{1}{\sqrt{2}}\left(\phi_1^{\prime}+i\chi_1^{\prime}\right)\;,\quad
\varphi_2^{\prime 0} = \frac{1}{\sqrt{2}}\left(\phi_2^{\prime}+i\chi_2^{\prime}\right)\;,\quad
\end{align}
where the VEVs of these neutral components are given by 
$\langle \varphi_1^0\rangle = v_1^{}/\sqrt{2}$,
$\langle \varphi_2^0\rangle = v_2^{}/\sqrt{2}$, 
$\langle \varphi_1^{\prime 0}\rangle= 0$ and  
$\langle \varphi_2^{\prime 0}\rangle = 0$.  
%By definition, $\varphi_1^{\prime 0}$ and $\varphi_2^{\prime 0}$ do not receive VEVs.
Introducing
\begin{equation}
v=(\sqrt{2}G_F^{})^{-1/2}\simeq 246\;\text{GeV}\;, 
\end{equation}
and the mixing angle $\beta$, 
we express $v_1$ and $v_2$ as 
$v_1 \equiv v \cos\beta$ and
$v_2 \equiv v \sin\beta$.  
The vacuum conditions for the Higgs potential are given by 
\begin{align}
\frac{1}{v}\left.\frac{\partial V_H^{}}{\partial \phi_1^{}}\right|_{\phi_i^{}=0}=&
c_{\beta}\left((M_{1}^2)_{11}^{}+\frac{m_Z^2}{2}c_{2\beta}\right)
-s_{\beta}(M_{3}^2)_{11}^{}=0\;,
\displaybreak[0]\nonumber\\
\frac{1}{v}\left.\frac{\partial V_H^{}}{\partial \phi_2^{}}\right|_{\phi_i^{}=0}=&
s_{\beta}\left((M_{2}^2)_{11}^{}-\frac{m_Z^2}{2}c_{2\beta}\right)
-c_{\beta}(M_{3}^2)_{11}^{}=0\;,
\displaybreak[0]\nonumber\\
\frac{1}{v}\left.\frac{\partial V_H^{}}{\partial \phi_1^{\prime}}\right|_{\phi_i^{}=0}=&
c_{\beta}(M_{1}^2)_{12}^{}
-s_{\beta}(M_{3}^2)_{21}^{}=0\;,
\displaybreak[0]\nonumber\\
\frac{1}{v}\left.\frac{\partial V_H^{}}{\partial \phi_2^{\prime}}\right|_{\phi_i^{}=0}=&
s_{\beta}(M_{2}^2)_{12}^{}
-c_{\beta}(M_{3}^2)_{12}^{}=0\;.
\end{align}
Solving this set of conditions, one can eliminate
$(M_1^2)_{11}^{}$, $(M_2^2)_{11}^{}$, $(M_1^2)_{12}^{}$, and $(M_2^2)_{12}^{}$.
%The formulae for the mass matrices in this paper are written in this manner.

%\subsection{Mass matrix}

After imposing the vacuum conditions,
the mass matrices $M_A^2$, $M_{H^\pm}^2$ and
$M_H^2$ for the CP-odd, charged and
CP-even scalar component states are
respectively obtained in the basis of
$(\Phi_1, \Phi_2, \Phi^{\prime}_1, \Phi^{\prime}_2)$.    
It is however more useful to work
the mass matrices of the CP-odd scalar bosons
and the charged Higgs bosons in the 
gauge eigenstate basis (the so-called Georgi basis)  as~\cite{Georgi}
\begin{align}
\bar{M}_A^2=&O_0^TM_A^2O_0
=
 \begin{pmatrix}
0&0&0&0\\
0&
\frac{2(M_3^2)_{11}}{s_{2\beta}}&
\frac{(M_3^2)_{21}}{c_{\beta}}&
\frac{(M_3^2)_{12}}{s_{\beta}}\\
0&
\frac{(M_3^2)_{21}}{c_{\beta}}&
(M_1^2)_{22}^{}+\frac{m_Z^2}{2}c_{2\beta}&
(M_3^2)_{22}^{}\\
0&
\frac{(M_3^2)_{12}}{s_{\beta}}&
(M_3^2)_{22}^{}&
(M_2^2)_{22}^{}-\frac{m_Z^2}{2}c_{2\beta}
  \end{pmatrix}\;, 
\end{align}
\begin{align}
\bar{M}_{H^{\pm}}^2=&O_0^TM_{H^{\pm}}^2O_0
=\bar{M}_A^2
+m_W^2\begin{pmatrix}
0&0&0&0\\
0&1&0&0\\
0&0&-c_{2\beta}&0\\
0&0&0&c_{2\beta}
\end{pmatrix}\;,
\end{align}
with the orthogonal matrix  
\begin{equation}
O_0=\begin{pmatrix}
c_{\beta}&s_{\beta}&0&0\\
-s_{\beta}&c_{\beta}&0&0\\
0&0&1&0\\
0&0&0&1
\end{pmatrix}\;,
\end{equation}
where we used the abbreviation such
as $\sin\theta=s_\theta$ and
$\cos\theta=c_\theta$. 
In this basis the massless modes, which are Nambu-Goldstone bosons to be
absorbed by the longitudinal modes of the weak gauge bosons,
are separated in the mass matrices. 
The basis taken here is essentially the same as that discussed in Ref.~\cite{Gupta}.
It is also useful to rotate the mass matrix for the CP-even scalar bosons as 
\begin{align}
\bar{M}_H^2=&O_0M_H^2O_0^T\nonumber\\
=&
\begin{pmatrix}
m_Z^2c_{2\beta}^2&-m_Z^2s_{2\beta}c_{2\beta}&0&0\\
-m_Z^2s_{2\beta}c_{2\beta}&
m_Z^2s_{2\beta}^2+\frac{2(M_3^2)_{11}}{s_{2\beta}}&
\frac{-(M_3^2)_{21}}{c_{\beta}}&
\frac{(M_3^2)_{12}}{s_{\beta}}\\
0&
\frac{-(M_3^2)_{21}}{c_{\beta}}&
(M_1^2)_{22}^{}+\frac{m_Z^2}{2}c_{2\beta}&
-(M_3^2)_{22}\\
0&
\frac{(M_3^2)_{12}}{s_{\beta}}&
-(M_3^2)_{22}&
(M_2^2)_{22}^{}-\frac{m_Z^2}{2}c_{2\beta}
\end{pmatrix}\;.
\end{align}

\section{Decoupling property of the extra doublet fields}

\subsection{Nondecoupling effects in the large mass regime}

In general, new physics can be tested not only by direct searches
but also by indirect searches. The indirect searches are performed
by precise experiments to find effects of a heavy new physics particle
on the observables which are well predicted in the low energy
theory such as the SM. Such new particle effects on the low
energy observables usually decouple in the large mass limit after
the renormalization calculation is completed.
This is known as the decoupling theorem proposed by Appelquist
and Carazzone~\cite{dec-theorem}. 
It is also known that the decoupling theorem does not hold when the new
particles receive their masses from the VEV of the Higgs boson.
In fact, there is a class of the new physics models where nondecoupling effects of
heavy particles can appear on the low energy observables.
For example, chiral fermions such as
quarks and charged leptons cannot have the mass term because of the
chiral symmetry, so that their masses are generated after the chiral
symmetry is spontaneously broken by the VEV.
Therefore, the effect of a heavy chiral fermion does not decouple, and
it appears as powerlike or logarithmic contributions of the mass in the
predictions for the low energy observables. 
Another example is the additional scalar fields in extended Higgs
sectors 
whose masses are typically expressed as
\begin{align}
 m_\varphi^2 \sim  \tilde{M}^2 + \lambda^\prime v^2, 
\end{align}
where $\lambda^\prime$ is the coupling constant with the SM-like Higgs boson 
and $\tilde{M}$ is the invariant mass parameter which is irrelevant to the
VEV. When $\lambda^\prime v^2 \gsim \tilde{M}^2$, the situation is similar to the case
of chiral fermions, so that the effect of these scalar bosons do not
decouple in the large $m_\varphi$ regime. 

The indirect effects of these nondecoupling particles appear in the low
energy observables at the tree level, or at loop levels such as the electroweak $S$,
$T$ and $U$ parameters~\cite{peskin-takeuchi} and 
vertex corrections to the SM coupling constants 
like the $\gamma\gamma h$ vertex~\cite{ggh} and the $hhh$ vertex~\cite{hhh4g,KOSY}. 
In particular, the nondecoupling effects of chiral
fermions and additional scalar bosons to the triple Higgs boson
coupling are known to give quartic powerlike
contributions for the heavy particle
masses in the radiative corrections, so that  
the effect can be very significant.
As shown in Ref.~\cite{KOSY}, in the two Higgs doublet model, the deviation
from the SM prediction can be of
order 100 \% without contradiction with
perturbative unitarity~\cite{pu2}. 
This nondecoupling
effect on the triple Higgs boson
coupling would be applied to realize
the strong first-order phase
transition~\cite{ewbg-thdm2}
which is required for successful electroweak
baryogenesis~\cite{B-EW,B-EW2,B-EW3}.
%x
It is obvious that an excessive 
nondecoupling effect is bounded by
the theoretical constraints such as
perturbative unitarity~\cite{pu} and triviality~\cite{triv},
because the large mass of new particles
directly means a large coupling constant.

Let us consider the effect of the
heavy particles in  
supersymmetric standard models. 
In general, a  SUSY Higgs potential is
composed of the D-term, the F-term
and the soft-breaking term;
\begin{align}
%V = |D|^2 + |F|^2 + ({\rm soft\;
% breaking\; term}). 
V = |D|^2 + |F|^2 + (\mbox{soft-breaking term}). 
\end{align}
Quartic coupling constants in the
potential can come from both the
D-term and the F-term.  
In the MSSM, however, because of the multi-doublet structure
only D-terms contribute to them, 
which are given by gauge coupling constants.  
Consequently, the mass of the lightest 
CP-even Higgs boson is determined by the gauge
coupling constants and the VEVs at the tree level, which is less than $m_Z$.
A substantial F-term contribution enters into the Higgs potential
at the one-loop level via the superpotential
$W \simeq - y_t U_{R3}^c H_2 Q_{L3}  + ...$,
where $y_t$ ($ =  \sqrt{2}m_t/v\sin\beta \simeq (Y_u)_{33}$)
is the top Yukawa coupling constant. 
The corrected mass of $h$ is described by 
\begin{align}
 m_h^2 \sim m_Z^2 \cos^22\beta + \frac{3}{2\pi^2} \frac{m_t^4}{v^2} \ln
 \frac{m_{\rm stop}^2}{m_t^2}.
\end{align}
This one-loop correction shows a nondecoupling property in the large
mass limit of stops. Consequently $m_h$ can be
above the LEP bound at least when one of the stops is heavy
enough.
There are also contributions from the $\mu$ parameter
and the soft-breaking $A_t$ ($A_b$) parameter when there is the left-right
mixing in the stop (sbottom) sector. Their one-loop effects can also be
nondecoupling and then can be significant to some extent 
when they are taken to be as large as the scale of the
soft-SUSY-breaking mass $m_{\rm SUSY}$.
In the NMSSM and the MSSM with triplets, $m_h$ can be 
significantly enhanced by the F-term
contribution~\cite{mh-NMSSM,mh-NMSSM2,mh-triplet}.
Notice that these F-term contributions should vanish in the
SUSY limit due to the nonrenormalization theorem.    
These F-term contributions also affect the prediction
on the other SM observables such as the triple Higgs boson coupling constants
 $\lambda_{hhh}$ similarly to the case of non-SUSY extended Higgs models
which we already discussed above~\cite{ksy}.

%[Models which reduce into the MSSM]
On the other hand, a typical example for extended SUSY Higgs sectors
without interactions from the tree-level F-term is that 
with only multi-doublet structures, such as the 4HDSSM.
In this class of models, if there is no mixing between
the light two doublet fields and the additional ones, 
the effects of the extra fields on the MSSM observables become
suppressed due to the decoupling theorem
when the extra doublet fields are heavy, and the model
behaves like the MSSM. 
However, nonvanishing effects can appear through the B-term
mixing when $(M_3^2)_{12}$ or $(M_3^2)_{21}$
grows with $(M_1^2)_{22}$ or $(M_2^2)_{22}$. 
These effects appear at the tree level, so that
they would give substantial modifications in the predictions
in the MSSM for the low energy observables.
We stress that these nonvanishing effects due to the B-term mixing 
are not the nondecoupling effects which appear in the large mass limit
for the new particles when $\lambda' v^2 \gsim \tilde{M}^2$ as a
consequence of violation of the decoupling theorem.
In this sense, we call the nonvanishing B-term mixing effect
as the quasi-nondecoupling effect.
Notice that the quasi-nondecoupling effect only appears
in the predicted values in the MSSM.
It gives modifications in the MSSM predictions  
such as the masses of $h$, $H$ and $H^\pm$
and the mixing angle $\alpha$
as well as coupling constants for the
MSSM-like Higgs bosons. Such an effect, however, does disappear
in the predictable SM coupling constants
of $h\gamma\gamma$, $hWW$, $hZZ$ and $hhh$ 
in the SM-like limit ($m_A \to \infty$)
according to the decoupling theorem.

Therefore, we would like to address the question of how the extra
doublet fields in the extended SUSY model can affect the
observables which appear in the MSSM, such as  
the mass $m_{\phi}$ ($\phi$ represent $h$, $H$ and $H^\pm$),   
the mixing angle $\alpha$, the vertex $F_{\phi'VV}$ ($V=W^\pm$ and 
$Z$; $\phi'=h$ and $H$)
as well as quarks and leptons $Y_{\phi''ff'}$ ($f,f'$
represent fermions and $\phi''=h$, $H$, $A$ and $H^\pm$).
Deviations from the renormalized MSSM observable parameters may
be expressed as 
\begin{align}
 m_{\phi} &\simeq m_{\phi}^{\rm
 MSSM} \left(1+
 \delta_{\phi} \right) \;, \\
 \sin^2(\beta-\alpha_{\rm eff}) &\simeq
 [\sin^2(\beta-\alpha)]^{\rm MSSM}
 \left(1 + \delta_s \right) \;, \label{eq:delta_s}\\
 F_{\phi' VV} &\simeq F_{\phi' VV}^{\rm
 MSSM} \left(1 + \delta_{\phi' VV} \right) \;, \\
 Y_{\phi'' ff'} &\simeq Y_{\phi'' ff'}^{\rm
 MSSM} \left(1 + \delta_{\phi'' ff'}\right)\;, 
\end{align}
where $\delta_\phi$, $\delta_s$, $\delta_{\phi'VV}$ or
$\delta_{\phi''ff'}$ represent the effect of the extra heavy scalar
fields on each observable in the extended SUSY models.
In this paper, we study 
$\delta_h$, $\delta_H$, $\delta_{H^\pm}$ and $\delta_s$ in the 4HDSSM.
The MSSM predictions are evaluated  at the one-loop level using
the on-shell renormalization scheme in Ref.~\cite{dabelstein}. 
 We do not discuss the effect on  the Yukawa coupling constants in
 this paper, which will be studied in details elsewhere~\cite{aksy-flavor}.

\subsection{Definition of the large mass
  limit and the decoupling property in the 4HDSSM}

 The soft-breaking mass parameters $(M_3^2)_{ij}$
 come from the B-terms in Eq.~(\ref{soft-term}).
When we consider the case with $(M_3^2)_{12}=(M_3^2)_{21}=0$, the
mass matrices $\bar{M}_A^2$, $\bar{M}_{H^\pm}^2$ and  $\bar{M}_H^2$
are block diagonal. The upper $2\times 2$ submatrix in each mass matrix
corresponds to that in the MSSM; i.e.,
$2(M_3^2)_{11}/s_{2\beta} \to m_A^2$, 
and the other $2\times 2$ submatrix corresponds to that for
the extra two scalar bosons.
They are separated completely in this case.
The model effectively becomes the MSSM
in the large mass limit of the extra scalar bosons.    
On the other hand, in the case with nonzero $(M_3^2)_{12}$ or $(M_3^2)_{21}$,
the masses of the light scalars $h$, $H$ and $H^\pm$  are modified 
from the MSSM predictions by the mixing via the B-terms
between $\Phi_1$ and $\Phi_1'$ or between $\Phi_2$ and $\Phi_2'$.
These effects are expected to be nonvanishing when
$(M_3^2)_{12}$ or $(M_3^2)_{21}$
grows with taking a similar value to
the 3-3 or 4-4 component in the mass matrices such as
$(M_1^2)_{22}$ or $(M_2^2)_{22}$. 
We here discuss these effects in
details in the following. 

We start from discussing the CP-odd scalar mass matrix. 
In order to examine the decoupling property of the mass matrices, we
further rotate $\bar{M}_A^2$ as  
\begin{equation}
\hat{M}_A^2=
\begin{pmatrix}
1&0&0&0\\
0&1&0&0\\
0&0&c_{\bar{\theta}}&s_{\bar{\theta}}\\
0&0&-s_{\bar{\theta}}&c_{\bar{\theta}}\\
\end{pmatrix}
\bar{M}_A^2
\begin{pmatrix}
1&0&0&0\\
0&1&0&0\\
0&0&c_{\bar{\theta}}&-s_{\bar{\theta}}\\
0&0&s_{\bar{\theta}}&c_{\bar{\theta}}\\
\end{pmatrix}
=
\begin{pmatrix}
0&0&0&0\\
0&\frac{2(M_3^2)_{11}^{}}{s_{2\beta}}&k^{\prime}M^2&kM^2\\
0&k^{\prime}M^2&M^2&0\\
0&kM^2&0&rM^2
\end{pmatrix}\;, 
\end{equation} 
where 
\begin{equation}
\tan2\bar{\theta}=\frac{2(M_3^2)_{22}}{(M_1^2)_{22}-(M_2^2)_{22}+m_Z^2c_{2\beta}}\;, 
\end{equation}
and $M^2$, $k$,
$k'$ and $r$ are defined
such that 
\begin{align}
&(M_3^2)_{21}=\kappa_{21}M^2\;,\quad
(M_3^2)_{12}=\kappa_{12}M^2\;,\quad\nonumber\\
&
(M_1^2)_{22}c_{\bar{\theta}}^2+
(M_2^2)_{22}s_{\bar{\theta}}^2
 +(M_3^2)_{22} s_{2\bar{\theta}}
+\frac{m_Z^2}{2}c_{2\beta}c_{2\bar{\theta}}=M^2\;,\quad\nonumber\\
&(M_1^2)_{22}s_{\bar{\theta}}^2+
(M_2^2)_{22}c_{\bar{\theta}}^2
-(M_3^2)_{22} s_{2\bar{\theta}}
-\frac{m_Z^2}{2}c_{2\beta}c_{2\bar{\theta}}=rM^2\;, 
\end{align}
and 
\begin{align}
k^{\prime}=\frac{c_{\bar{\theta}}}{c_{\beta}}\kappa_{21}+\frac{s_{\bar{\theta}}}{s_{\beta}}\kappa_{12}\;,\quad
k=-\frac{s_{\bar{\theta}}}{c_{\beta}}\kappa_{21}+\frac{c_{\bar{\theta}}}{s_{\beta}}\kappa_{12}\;.
\end{align}
These parameters are relevant to the extra doublets, then 
the decoupling limit is taken as  $M^2\to \infty$.
Here we assume that $m_A^2\ll M^2$ and we treat $m_A^2/M^2$ as an 
expansion parameter.
One of the eigenvalues of $\bar{M}_A^2$ should be $m_A^2$,
the mass of the lightest CP-odd
Higgs boson $A$, which should coincide
with $2(M_3^2)_{11}/s_{2\beta}$ in
the limit of $M \to \infty$ if $k=k'=0$. 
In generic cases, after diagonalizing the mass matrix,  
$(M_3^2)_{11}$ is expressed in terms of $m_A^2$ as 
\begin{equation}
\frac{2(M_3^2)_{11}}{s_{2\beta}}=
m_A^2\left\{ \frac{k^2+(1+k^{\prime 2})r^2}{r^2}+\mathcal{O}\left(\frac{m_A^2}{M^2}\right)\right\}
+M^2\frac{k^2+rk^{\prime 2}}{r}
\;.
\end{equation} 
The mass eigenvalues for heavier
states $A_1$ and $A_2$ are 
\begin{equation}
m_{A_1}^2\simeq a_1M^2\;,\quad
m_{A_2}^2\simeq a_2M^2\;,
\end{equation}
where $a_1$ and $a_2$ are given by 
\begin{align}
a_1=&\frac{k^2+rk^{\prime 2}+r(1+r)
-\sqrt{\left\{k^2+rk^{\prime 2}-r(r-1)\right\}^2+4k^2r(r-1)}}{2r}\;,\nonumber\\
a_2=&\frac{k^2+rk^{\prime 2}+r(1+r)
+\sqrt{\left\{k^2+rk^{\prime 2}-r(r-1)\right\}^2+4k^2r(r-1)}}{2r}\;.
\end{align}
We note that $a_1 \to 1$ and $a_2
\to r$ for $k=k' \to 0$. 

\begin{figure}\label{fig:del_ch}
\begin{center}
\begin{tabular}{cc}
\includegraphics{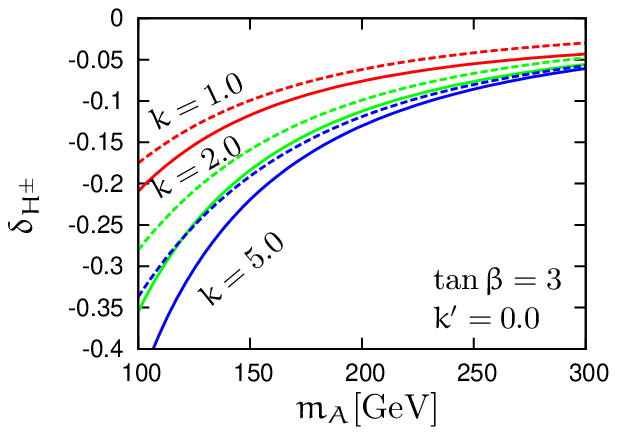}&
\includegraphics{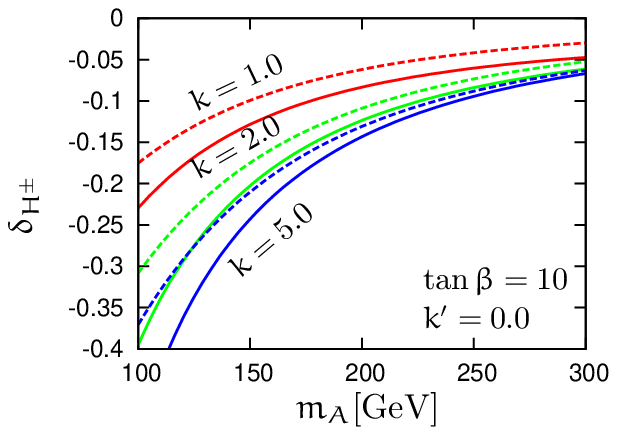}\\
\includegraphics{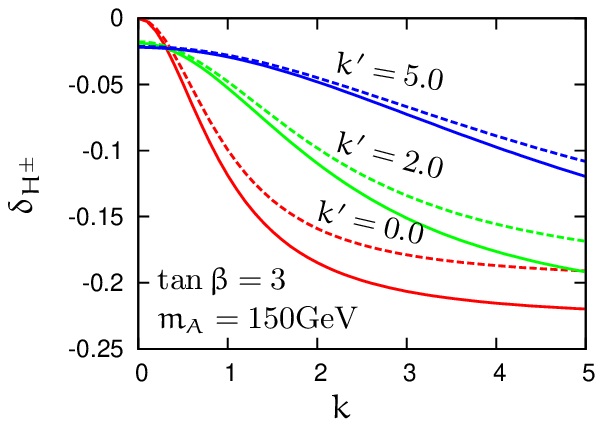}&
\includegraphics{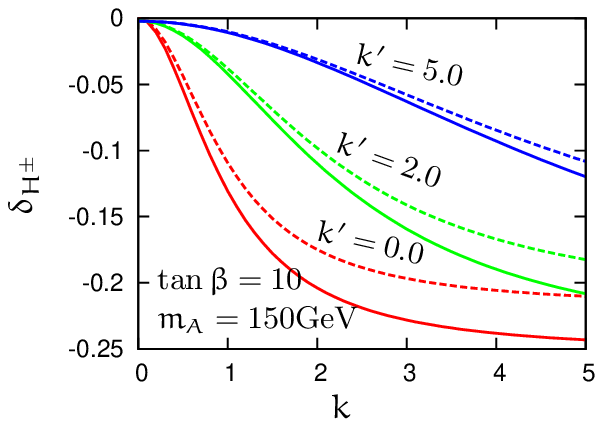}
\end{tabular}
\end{center}
\caption{The deviation $\delta_{H^\pm}$ defined in Eq.~(\ref{eq:delta})
 due to the quasi-nondecoupling effect of the B-term mixing
 parameterized by $k$ and $k'$ in the 4HDSSM.
 We here take $M=500$ GeV, $r=1$ and $\bar{\theta}=0$.
 The SUSY breaking scale for the MSSM particles is taken to be 1 TeV,
 and the trilinear soft-breaking parameters $A_t$ and $A_b$
 as well as the $\mu$ parameter are taken to be zero.
The upper figures: $\delta_{H^\pm}$ as a function of $m_A$ for $\tan\beta=3$
 (left) and $\tan\beta=10$ (right) for $k=1.0$, $2.0$ and $5.0$ with fixed $k'(=0.0)$. 
The lower figures: $\delta_{H^\pm}$ as a function of $k$ for $\tan\beta=3$
 (left) and $\tan\beta=10$ (right) for $k'=0.0$, $2.0$ and $5.0$
 with the fixed $m_A$ (= 150 GeV).
In all figures, the solid curves
 are the results from the full numerical calculation, while the dotted
 curves are those by using the approximated formula in Eq.~(\ref{delmch_approx}). 
 }
\end{figure}

For the charged Higgs mass matrix, via
the similar procedure to the case of
the CP-odd Higgs bosons, we obtain
the deviation in $m_{H^\pm}$,  
the mass eigenvalue for the lightest
charged scalar $H^\pm$, from
the MSSM prediction as 
\begin{equation}
m_{H^{\pm}}=\sqrt{(m_{H^{\pm}}^2)^{\rm MSSM}}(1+\delta_{H^\pm}), \label{eq:delta}
\end{equation}
where 
\begin{equation}
\delta_{H^\pm} = 
-\frac{1}{2}\frac{m_W^2}{m_A^2+m_W^2}
\frac{k^2+k^{\prime 2} r^2
-c_{2\beta}\left\{
	(k^2-k^{\prime 2}r^2)c_{2\bar{\theta}}
	+2kk^{\prime}rs_{2\bar{\theta}}
	\right\}}
{\left\{k^2+(1+k^{\prime 2})r^2\right\}}
+\mathcal{O}\left(\frac{m_A^2}{M^2}\right)\;, \label{delmch_approx}
\end{equation}
and $(m_{H^{\pm}}^2)^{\rm MSSM}$ is 
the prediction in the MSSM renormalized in the on-shell
scheme~\cite{dabelstein}, which is simply given by~\cite{CKY}  
\begin{equation}
(m_{H^{\pm}}^2)^{\rm MSSM}=m_A^2+m_W^2
 - \Pi_{H^+H^-}^{\rm 1PI}(m_A^2+m_W^2)
 + \Pi_{AA}^{\rm 1PI}(m_A^2)
 + \Pi_{WW}^{\rm 1PI}(m_W^2), 
\end{equation}
where $\Pi_{\phi\phi}^{\rm 1PI}(p^2)$
represent the one particle irreducible diagram contributions
to the two point function of the field $\phi$ at the squared
momentum $p^2$.
Masses of the heavier charged scalar
bosons $H_1^\pm$ and $H_2^\pm$ are obtained as 
\begin{equation}
m_{H_1^{\pm}}^2\simeq a_1M^2\;,\quad
m_{H_2^{\pm}}^2\simeq a_2M^2\;.
\end{equation}

In Fig.~1, we show the numerical results for the deviation $\delta_{H^\pm}$ defined 
in Eq.~(\ref{eq:delta}) due to the quasi-nondecoupling
effect of the B-term mixing parameterized by $k$ and $k'$ in our model.
The solid curves in the figures 
represent the results from the full numerical calculation, while the dotted
 curves are those by using the approximated formula in
 Eq.~(\ref{delmch_approx}).
The deviation $\delta_{H^\pm}$ turns out to be negative, and amounts
to $-20$ \% for a relatively small value of $m_A$. The magnitude of
 the deviation is smaller for a larger value of $m_A$, but still a few times
$-1$ \% even for $m_A=300$ GeV.
On the other hand, the deviation is not very sensitive to $\tan\beta$.
We note that the results
are insensitive to the details of the MSSM parameters such as the soft-breaking mass
parameters, the $\mu$ parameter and the trilinear $A_{t,b}$ parameters.
In fact, when $\mu$ and $A_{t,b}$ are varied in the phenomenologically
acceptable regions, the radiative corrections vary at most
from $-2$ \% to $+2$ \%. We have confirmed that our results on 
the one-loop correction in the MSSM agree with those given in
Ref.~\cite{CKY}. 
The mass of $H^\pm$ can be determined with the accuracy of
a few percent via the decays of $H^\pm \to \tau \nu $ and $H^\pm \to tb$ at
the LHC~\cite{assamagan}, and with the statistical 
error of less than 1 \% via $e^+e^- \to H^+H^-$ at the ILC~\cite{djouadi}.
The mass of $A$ can also be determined with the resolution about 2\%
via the decays $A \to \mu^+\mu^-$ at the LHC, while at the ILC
it can be measured with the precision 0.2~\%
via $e^+e^- \to HA$~\cite{djouadi}.
Therefore, the quasi-nondecoupling effect on $m_{H^\pm}$ can be
extracted when both $m_{H^\pm}$ and $m_{A}$ are
measured at future collider experiments.
The prediction on $m_{H^\pm}$ (not on $\delta_{H^\pm}$) in the 4HDSSM is
 shown in Fig.~5 with the comparison of the result in the MSSM.

Next, the CP-even scalar mass matrix $\bar{M}_H^2$ can also be
diagonalized. We first define $\hat{M}_H^2$ by  
\begin{align}
\hat{M}_H^2=
\begin{pmatrix}
1&0&0&0\\
0&1&0&0\\
0&0&c_{\bar{\theta}}&-s_{\bar{\theta}}\\
0&0&s_{\bar{\theta}}&c_{\bar{\theta}}\\
\end{pmatrix}
\bar{M}_H^2
\begin{pmatrix}
1&0&0&0\\
0&1&0&0\\
0&0&c_{\bar{\theta}}&s_{\bar{\theta}}\\
0&0&-s_{\bar{\theta}}&c_{\bar{\theta}}\\
\end{pmatrix}\; , 
\end{align}
and  according to the usual mathematical procedure
$\hat{M}_H^2$ can be block-diagonalized 
 by rotating the basis with an appropriate orthogonal matrix $O_{MH}$ as    
\begin{align}
O_{MH}^T\hat{M}_H^2O_{MH}=&
\begin{pmatrix}
m_Z^2c_{2\beta}^2&-m_Z^2c_{2\beta}s_{2\beta}R&0&0\\
-m_Z^2c_{2\beta}s_{2\beta}R&m_A^2+m_Z^2s_{2\beta}^2R^2&
0&0\\
0&0&a_1M^2&0\\
0&0&0&a_2M^2
\end{pmatrix}%\nonumber\\
+\mathcal{O}\left(\frac{m_A^2}{M^2}\right)\;, \label{cpemm}
\end{align}
where $R$ is defined as 
\begin{equation}
R=\frac{1}{\sqrt{1+\frac{k^2}{r^2}+k^{\prime 2}}}\;.
\end{equation}
The upper $2 \times 2$ submatrix coincides to the mass matrix of
the two light scalar bosons $H$
and $h$ of the MSSM when $M\to \infty$ if $k=k'=0$. 
For the case with nonzero $k$ and $k'$, after diagonalizing the
$2\times 2$ submatrix by the mixing
angle $\alpha_{\rm eff}$ the mass eigenvalues of the CP-even
Higgs bosons are obtained as
\begin{align}
m_h^2=&
\frac{1}{2}\left[ m_A^2+m_Z^2\left(c_{2\beta}^2+R^2s_{2\beta}^2\right)
-\sqrt{\left\{(m_A^2-m_Z^2\left(1-(1-R^2)s_{2\beta}^2\right)\right\}^2
+4m_A^2m_Z^2s_{2\beta}^2R^2} \right. \nonumber\\
&\left. +m_A^2\mathcal{O}\left(\frac{m_A^2}{M^2}\right)+ \Delta_h^{\rm loop}
\right] 
\;,\nonumber\\
m_H^2=&
\frac{1}{2} \left[ m_A^2+m_Z^2\left(c_{2\beta}^2+R^2s_{2\beta}^2\right)
 +\sqrt{ \left\{m_A^2-m_Z^2\left(1-(1-R^2)s_{2\beta}^2\right)\right\}^2
 +4m_A^2m_Z^2s_{2\beta}^2R^2} \right. \nonumber\\
&\left. +m_A^2\mathcal{O}\left(\frac{m_A^2}{M^2}\right)
+ \Delta_H^{\rm loop}
\right] \;,  
\label{mh}
\end{align}
where $\Delta_h^{\rm loop}$ and $\Delta_H^{\rm loop}$ represent
the one-loop corrections in the MSSM.
The masses of heavier states
$H_1'$ and $H_2'$ are given by  
$m_{H_1'}^2 \simeq a_1M^2\left\{ 1+\mathcal{O}(m_A^2/M^2) \right\}$
and
$m_{H_2'}^2 \simeq a_2M^2\left\{ 1+\mathcal{O}(m_A^2/M^2) \right\}$.
The mixing angle $\alpha_{\rm eff}$
satisfies the relation  
\begin{equation}
\tan(\beta-\alpha_{\text{eff}})
=\frac{m_h^2-m_A^2-m_Z^2s_{2\beta}^2R^2}{m_Z^2c_{2\beta}s_{2\beta}R}
\left\{ 1+\mathcal{O}\left(\frac{m_A^2}{M^2}\right)
 + \Delta_{\tan(\beta-\alpha)}^{\rm loop}\right\}\;,  \label{tanbaeff} 
\end{equation}
where $\Delta_{\tan(\beta-\alpha)}^{\rm loop}$ is the one-loop
correction in the MSSM.
Notice that $m_h$ and $m_H$ given in Eq.~(\ref{mh}) and
$\tan(\beta-\alpha_{\rm eff})$ in Eq.~(\ref{tanbaeff}) do not depend on
the sign of $k$ and $k'$. 
\begin{figure}
\begin{center}
\begin{tabular}{cc}
\includegraphics{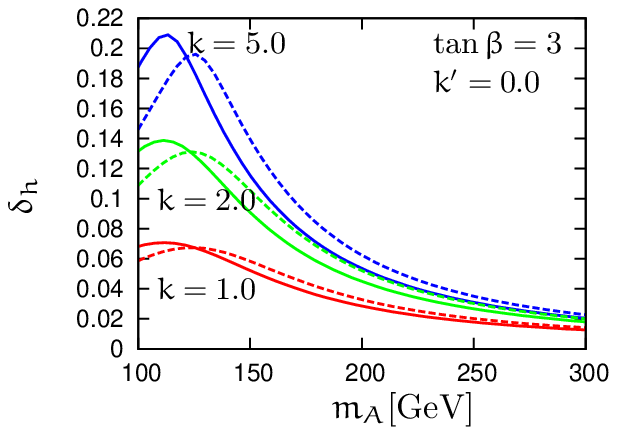}&
\includegraphics{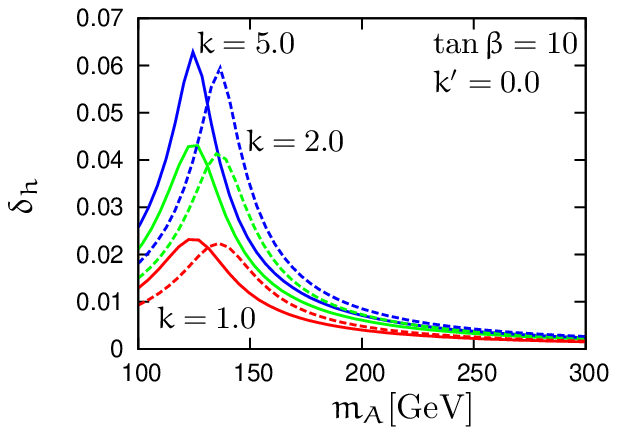} \\
\includegraphics{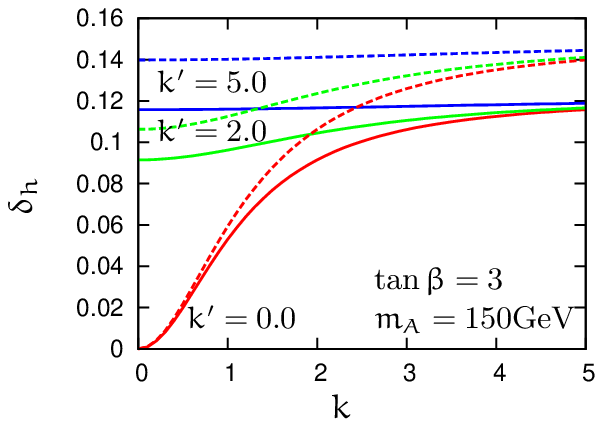}&
\includegraphics{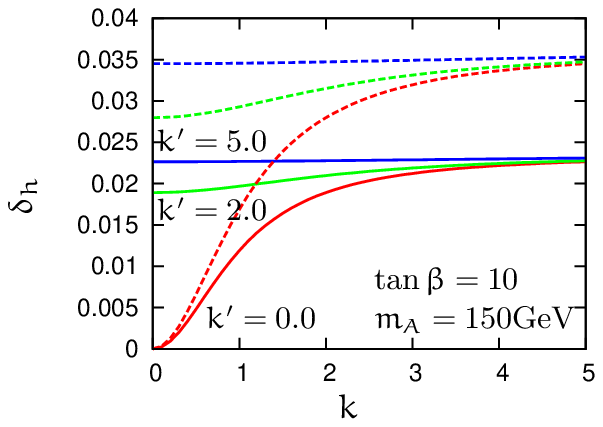}
\end{tabular}
\end{center}
\caption{The deviation $\delta_h$ in
 $m_h=m_h^{\text{MSSM}}(1+\delta_h)$ 
 due to the quasi-nondecoupling
 effect of the B-term mixing parameterized by $k$ and $k'$
  in the 4HDSSM, where  
 $m_h^{\text{MSSM}}$ is the renormalized mass of $h$.
  We here take $M=500$ GeV, $r=1$ and $\bar{\theta}=0$.
 The SUSY soft-breaking scale for the MSSM particles
 is taken to be 1TeV (solid curves) and 2 TeV (dotted curves),
 and the trilinear soft-breaking parameters $A_t$ and $A_b$ as well as 
 the $\mu$ parameter are taken to be zero.
The upper figures: $\delta_h$ as a function of $m_A$ for $\tan\beta=3$
 (left) and $\tan\beta=10$ (right) for $k=1.0$, $2.0$ and $5.0$ with fixed $k'(=0.0)$. 
The lower figures: $\delta_h$ as a function of $k$ for $\tan\beta=3$
 (left) and $\tan\beta=10$ (right) for $k'=0.0$, $2.0$ and $5.0$
 with the fixed $m_A$ (= 150 GeV). }\label{fig:del_h}
\end{figure}
\begin{figure}\label{fig:del_H}
\begin{center}
\begin{tabular}{cc}
\includegraphics{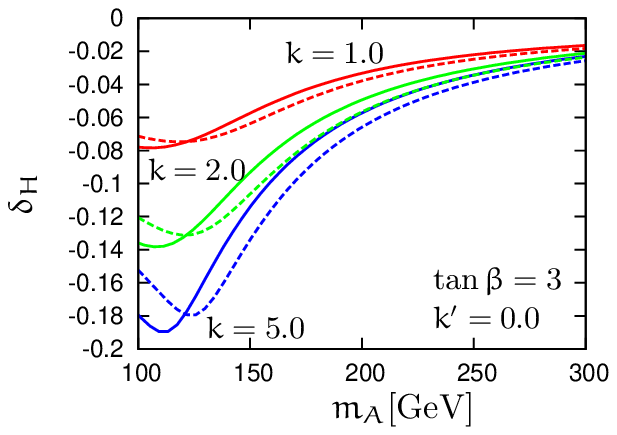}&
\includegraphics{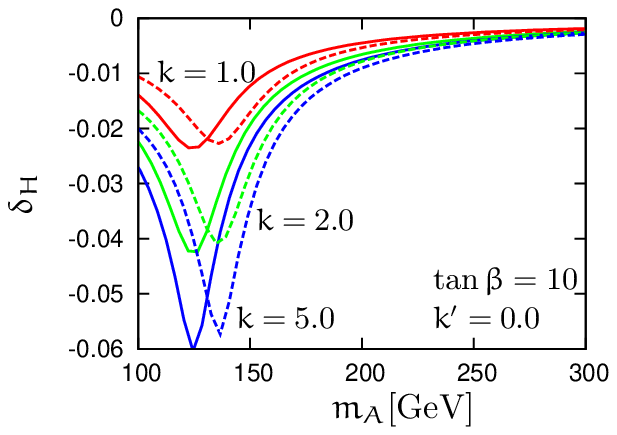}\\
\includegraphics{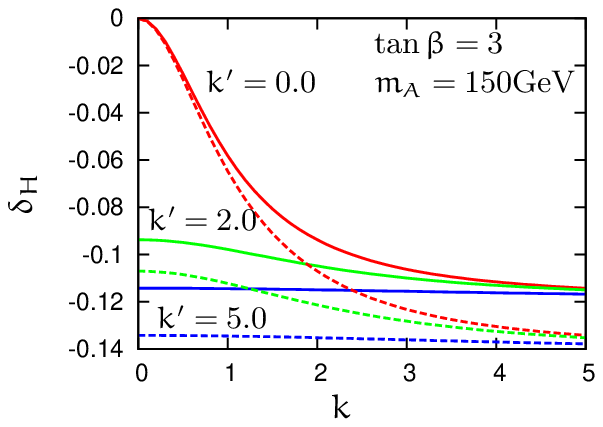}&
\includegraphics{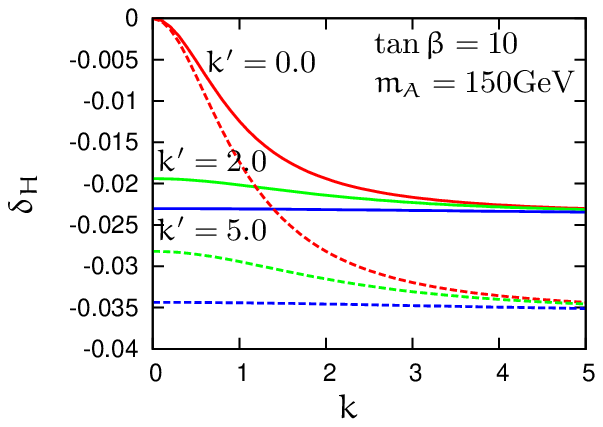}\\
\end{tabular}
\end{center}
 \caption{The deviation $\delta_H$ in $m_H=m_H^{\rm MSSM}(1+\delta_H)$ of
 the renormalized mass of the second lightest CP-even Higgs boson $H$
 due to the quasi-nondecoupling effect of the B-term mixing
 parameterized by $k$ and $k'$ in the 4HDSSM.
   We here take $M=500$ GeV, $r=1$ and $\bar{\theta}=0$.
The SUSY soft-breaking scale of the MSSM particles is taken to be 1 TeV
 (solid curves) and 2 TeV (dotted curves), and the trilinear
 soft-breaking parameters $A_t$ and $A_b$
 as well as the $\mu$ parameter are taken to be zero.
The upper figures: $\delta_H$ as a function of $m_A$ for $\tan\beta=3$
 (left) and $\tan\beta=10$ (right) for $k=1.0$, $2.0$ and $5.0$ with fixed $k'(=0.0)$. 
The lower figures: $\delta_H$ as a function of $k$ for $\tan\beta=3$
 (left) and $\tan\beta=10$ (right) for $k'=0.0$, $2.0$ and $5.0$
 with the fixed $m_A$ (= 150 GeV). } 
\end{figure}
\begin{figure}
\begin{center}
\begin{tabular}{cc}
\includegraphics{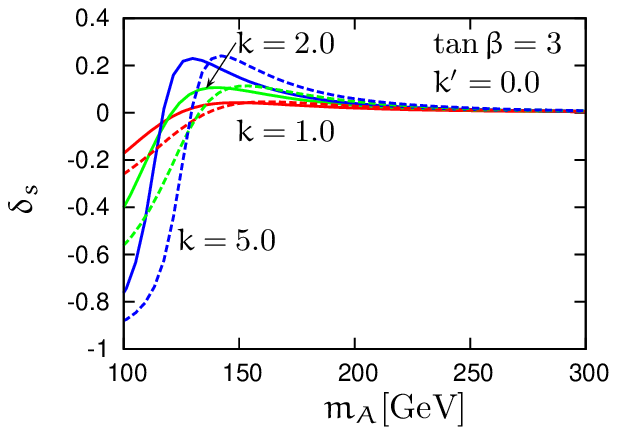}&
\includegraphics{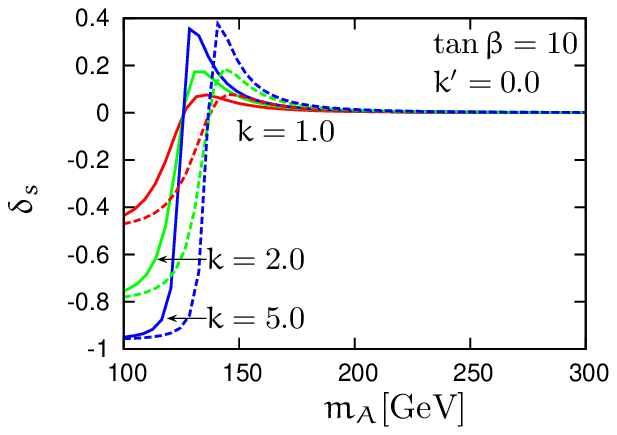}\\
\includegraphics{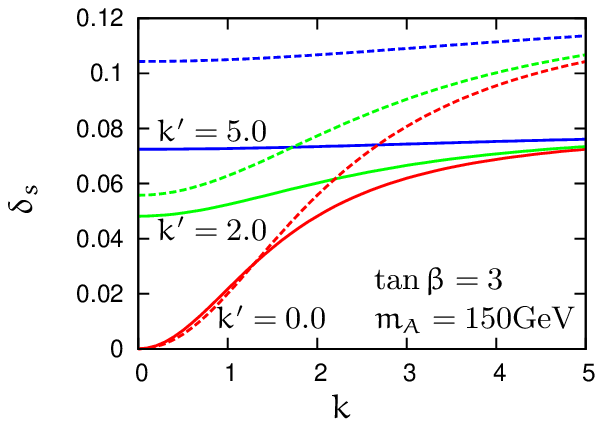}&
\includegraphics{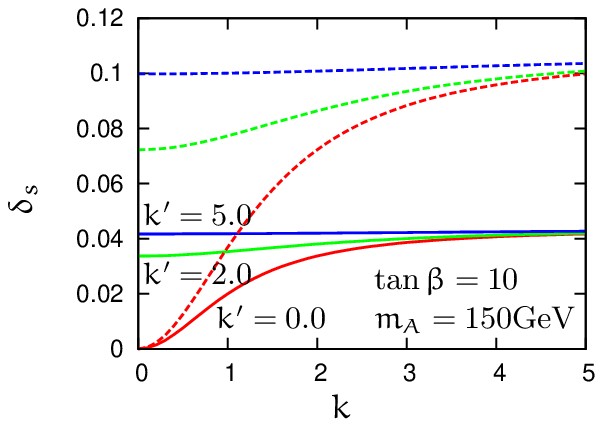}
\end{tabular}
\end{center}\label{fig:del_s}
 \caption{The deviation $\delta_s$ in Eq.~(\ref{eq:delta_s})
 due to the quasi-nondecoupling effect of extra doublet fields
 via the B-term mixing parameterized by $k$ and $k'$ in the 4HDSSM.
  We here take $M=500$ GeV, $r=1$ and $\bar{\theta}=0$.
 The SUSY soft-breaking scale of the MSSM particles is taken to be 1 TeV
 (solid curves) and 2 TeV (dotted curves), and the trilinear
 soft-breaking parameters $A_t$ and $A_b$ as well as
 the $\mu$ parameter are taken to be zero.
The upper figures: $\delta_s$ as a function of $m_A$ for $\tan\beta=3$
 (left) and $\tan\beta=10$ (right) for $k=1.0$, $2.0$ and $5.0$ with fixed $k'(=0.0)$. 
The lower figures: $\delta_s$ as a function of $k$ for $\tan\beta=3$
 (left) and $\tan\beta=10$ (right) for $k'=0.0$, $2.0$ and $5.0$
 with the fixed $m_A$ (= 150 GeV). }
\end{figure}
We note that the effective mixing angle
$\alpha_{\rm eff}$ contains information of the B-term quasi-nondecoupling
effects between $\Phi_1$ and $\Phi_1'$ or
between $\Phi_2$ and $\Phi_2'$ by $k$ and $k'$, but for
$m_A^2 \ll M^2$ the tree level
formula with the angle $\alpha$ in
the MSSM can still hold by replacing
$\alpha$ by $\alpha_{\rm eff}$ in a good approximation. For
example, the coupling constants of
the two light CP-even Higgs bosons with the
weak gauge bosons $V$ ($V=W^\pm$ and $Z^0$) in the case 
with nonzero $k$ and $k'$ are given
by 
\begin{align}
\Gamma_{VVh}^{}=-\frac{m_V^2}{v} \left\{ c_{\beta}(O_H)_{12}
 +s_{\beta}(O_H)_{22}\right\}
&=\frac{m_V^2}{v}
 \sin(\beta-\alpha_{\text{eff}})
 \left( 1+  \Delta_{hVV}^{\rm loop}\right)\;, \\
\Gamma_{VVH}^{}=-\frac{m_V^2}{v}
 \left\{ c_{\beta}(O_H)_{11}+s_{\beta}(O_H)_{21} \right\}
&=\frac{m_V^2}{v} \cos(\beta-\alpha_{\text{eff}})
\left( 1+ \Delta_{HVV}^{\rm loop}\right)\;,
\end{align}
where the matrix $O_H$ is given in Eq.~(B.5) in Appendix B, and
$\Delta_{hVV}^{\rm loop}$ and $\Delta_{HVV}^{\rm loop}$ represent 
radiative corrections in the MSSM.
Finally, in general, magnitudes of $k$ and $k'$ are not
necessarily smaller than 1, still it
is helpful to deduce the approximate
formulae assuming that they are
small;  
\begin{align}
m_h^2=&
(m_h^2)^{\rm MSSM} 
\left(
1+
\frac{m_Z^2s_{2\beta}^2(\frac{k^2}{r^2}+k^{\prime 2})}{\sqrt{(m_A^2-m_Z^2)^2+4m_Z^2m_A^2s_{2\beta}^2}}
+\mathcal{O}(k^4,k^{\prime 4}, k^2k^{\prime 2})
+\mathcal{O}\left(\frac{m_A^2}{M^2}\right)
\right)
\;,\\
m_H^2=&
(m_H^2)^{\rm MSSM} 
\left(
1-
\frac{m_Z^2s_{2\beta}^2(\frac{k^2}{r^2}+k^{\prime 2})}{\sqrt{(m_A^2-m_Z^2)^2+4m_Z^2m_A^2s_{2\beta}^2}}
+\mathcal{O}(k^4,k^{\prime 4}, k^2k^{\prime 2})
+\mathcal{O}\left(\frac{m_A^2}{M^2}\right)
\right) 
\;, \\
%\end{align}
%\begin{align}
\tan(\beta-\alpha_{\text{eff}})
=& [\tan(\beta-\alpha)]^{\rm MSSM} 
\left(
1+
\frac{(m_A^2-2m_h^2-m_Z^2s_{2\beta}^2)(\frac{k^2}{r^2}+k^{\prime 2})}
{2(m_A^2-m_h^2+m_Z^2s_{2\beta}^2)} 
+\mathcal{O}(k^4,k^{\prime 4}, k^2k^{\prime 2}) \right. \nonumber\\
&\left. \hspace{33mm}+\mathcal{O}\left(\frac{m_A^2}{M^2}\right)
\right)\;, 
\end{align}
where $(m_h^2)^{\rm MSSM}$, $(m_H^2)^{\rm MSSM}$ and
$[\tan(\beta-\alpha)]^{\rm MSSM}$ are the corresponding parameters
evaluated at the one-loop level assuming the MSSM.
In this paper, we have used the approximate one-loop formula given in
Ref.~\cite{dabelstein} in evaluating
 $(m_h^2)^{\rm MSSM}$, $(m_H^2)^{\rm MSSM}$ and
$[\tan(\beta-\alpha)]^{\rm MSSM}$.

In Fig.~2, we show the numerical results for
the deviation $\delta_h$ in $m_h=m_h^{\text{MSSM}}(1+\delta_h)$,
where $m_h^{\text{MSSM}}$ is the one-loop corrected mass
of $h$, due to the quasi-nondecoupling effect of the B-term mixing
parameterized by $k$ and $k'$ in the 4HDSSM.
The SUSY soft-breaking scale of the MSSM is taken to be 1 TeV and 2 TeV,
and the trilinear soft-breaking parameters $A_t$ and $A_b$ and the $\mu$
are taken to be zero.
It is found that $\delta_h$ is always positive.
This is understood from Eq.~(\ref{cpemm}).
The parameter $R$ is unity for $k=k'=0$, and is smaller
for larger values of $k$ and $k'$. A smaller value of $R$ ($R < 1$)  
reduces the value of the off-diagonal term in Eq.~(\ref{cpemm}), which
makes the mixing between the first two CP-even states weaker. Consequently,
the mass difference between $h$ and $H$ becomes smaller than the case
with the MSSM case with the same value of $m_A$ and $\tan\beta$.
The deviation takes its maximal 
values (6-20 \% for $\tan\beta =3$ and $2$-$5$ \% for $\tan\beta=10$)
around the crossing point ($m_A \sim 130$-$150$ GeV)
where the role of $h$ and $H$ are exchanged. 
For larger values of $m_A$ the magnitude of $\delta_h$ is smaller, but
it can be still 3-6 \% (about 1 \%) at $m_A=200$ GeV for $\tan\beta = 3$ $(10)$.
These values are substantial and can be tested by the precise
measurement of $m_h$ at the LHC (the ILC), where $m_h$ is expected to be
determined with about 0.1\% \cite{acc-mh-LHC} accuracy at the LHC, while 
at the ILC it is expected to be measured within 
less than 70 MeV \cite{acc-mh-ILC}) error.
The prediction on $m_{h}$ (not on $\delta_{h}$) in the 4HDSSM is 
shown in Fig.~5 with the comparison of the result in the MSSM.
We can see that in the 4HDSSM $m_{h}$ reaches its maximal value at a smaller $m_{A}$
than that in the MSSM, although the predicted upper bound on the $m_{h}$
is the same in both models.

In Fig.~3, we show 
the deviation $\delta_H$ in $m_H=m_H^{\text{MSSM}}(1+\delta_H)$,
where $m_H^{\text{MSSM}}$ is the one-loop corrected mass
of $H$, due to the quasi-nondecoupling effect of the B-term mixing
parameterized by $k$ and $k'$ in the 4HDSSM.
The SUSY parameters are taken as in the same way as Fig.~2.
As we discussed, the mixing of the light two CP-even states is
weakened by non-zero values of $k$ and $k'$, so that $m_H$
is smaller than the prediction in the MSSM. Therefore, 
$\delta_H$ is negative as we expect. The behavior of $\delta_H$
as a function of $m_A$ and $\tan\beta$ are similar to the case of
$\delta_h$ except for the sign. 
The magnitude is maximal around the crossing point ($m_A =130$-$150$
GeV), and amounts to $-18$ \% ($-5$ \%) for $\tan\beta=3$ (10).
At the LHC and the ILC, the mass of $H$ can be determined with the
similar precision to  that of $A$ mentioned in the previous paragraph.
The prediction on $m_{H}$ (not on $\delta_{H}$) in the 4HDSSM is 
shown in Fig.~5 with the comparison with the result in the MSSM.

In Fig.~4, we show the numerical results for
the deviation $\delta_s$ defined in Eq.~(\ref{eq:delta}),   
in which  $[\sin^2(\beta-\alpha)]^{\text{MSSM}}$ is the one-loop corrected
mixing factor $\sin^2(\beta-\alpha)$ evaluated in the MSSM. 
$\delta_s$ is the net deviation from the MSSM prediction 
due to the quasi-nondecoupling effect of the B-term mixing
parameterized by $k$ and $k'$ in the 4HDSSM.
The SUSY soft-breaking scale of the MSSM is taken to be 1 TeV and 2 TeV,
and the trilinear soft-breaking parameters $A_t$ and $A_b$ and the $\mu$
parameter are taken to be zero. 
In the figures, we can see that $\delta_s$ is negative when  
$m_A$ is smaller than the crossing point at $m_A \sim 130$-$150$ GeV,
while it is positive for larger $m_A$.  
The deviation can be as large as ${\mathcal O}(10)$ \% ($\tan\beta=3$)
and ${\mathcal O}(20)$ \% ($\tan\beta=10$) just above the crossing
point; i.e., at around 
$m_A \sim 140$-$150$ GeV.  
It is rapidly close to unity for larger values of $m_A$.
Notice that for larger soft-SUSY-breaking scale, a larger $\delta_s$
is possible. 
The prediction on $\sin^2(\beta-\alpha_{\rm eff})$ (not on
$\delta_{s}$)
in the 4HDSSM is shown in Fig.~5 with the comparison with the result in the MSSM.

\begin{figure}[t]
\begin{center}
\begin{tabular}{cc}
\includegraphics{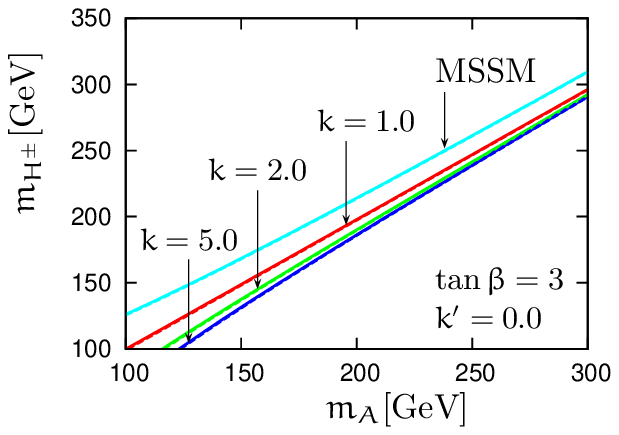}&
\includegraphics{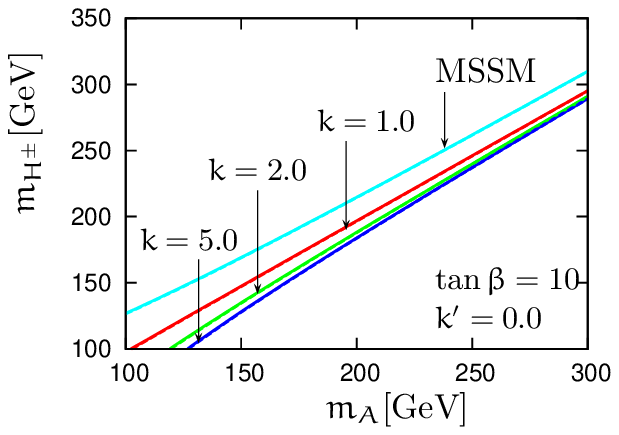}\\
\includegraphics{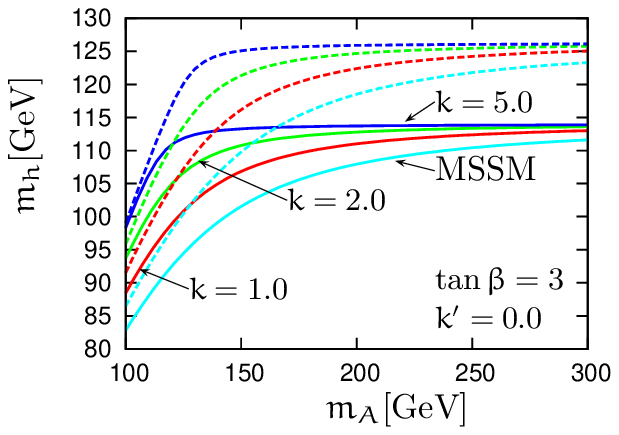}&
\includegraphics{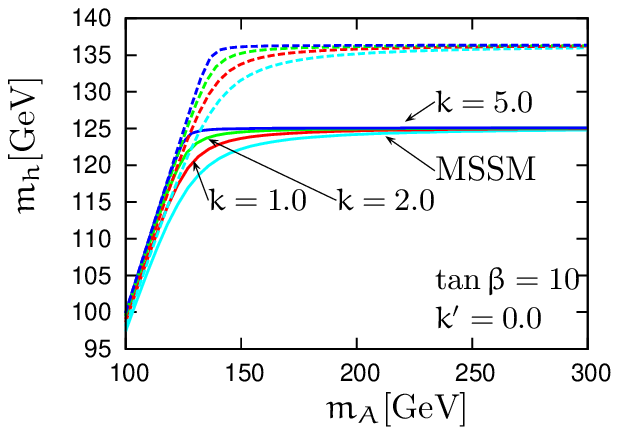}\\
\includegraphics{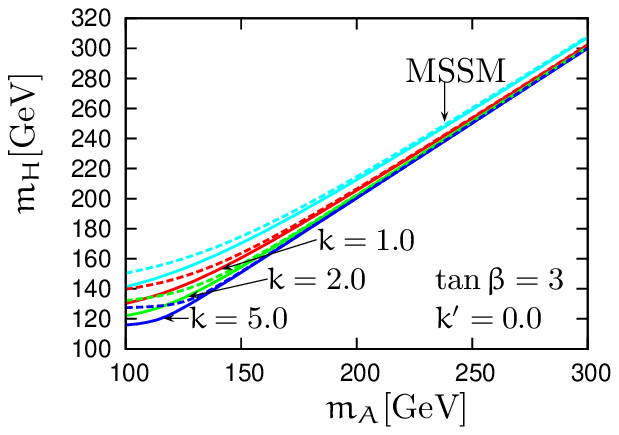}&
\includegraphics{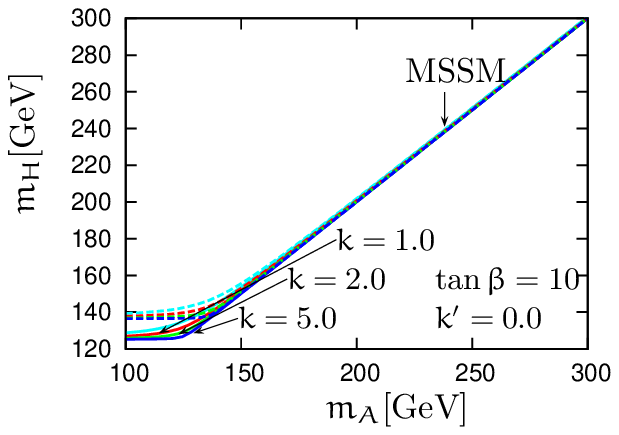}\\
\includegraphics{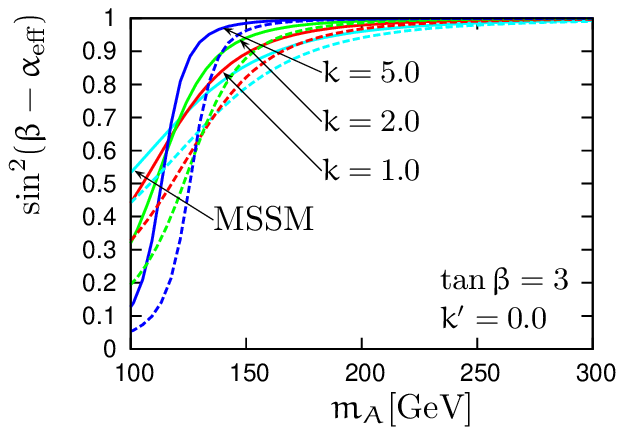}&
\includegraphics{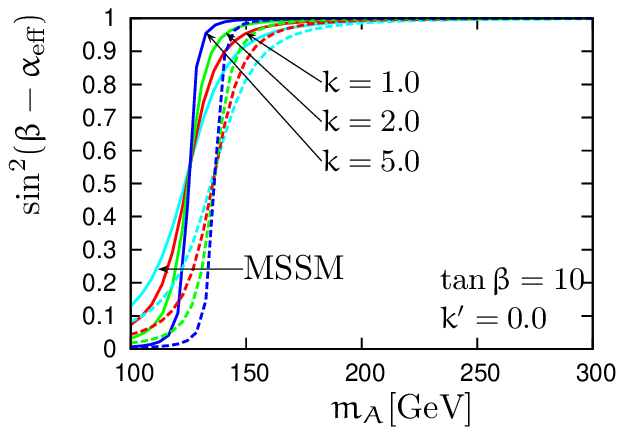}
\end{tabular}
\end{center}
\caption{The values of $m_{H^\pm}$,$m_{h}$,$m_{H}$ and
 $\sin^2(\beta-\alpha_{\rm eff})$ in the 4HDSSM and the MSSM 
 as a function of $m_A$ for $k=1.0$, $2.0$ and $5.0$.
 The soft-SUSY-breaking scale of the MSSM
 is set to be 1 TeV (solid curves) and 2 TeV (dotted curves).
 The trilinear soft-breaking parameters $A_t$ and $A_b$ as well as
 the $\mu$ parameter are taken to be zero.
 The other parameters are takes as 
 $M=500$ GeV, $r=1$, $\bar{\theta}=0$ and $k'=0$.
 Figures in the left column are for $\tan\beta=3$ and those in the right
 are for $\tan\beta=10$.}
\end{figure}

\section{Conclusions}

We have investigated the decoupling property of extra heavy doublet
fields in the 4HDSSM. 
Even without interaction terms from the tree-level F-term contribution
such as in the NMSSM, significant  
quasi-nondecoupling effects of extra scalar fields
can occur at the tree level due to the B-term mixing among the Higgs bosons.
We have deduced formulae for deviations in the MSSM observables 
in the decoupling region for the extra heavy fields. 
The possible modifications in the Higgs sector from the MSSM predictions 
have been studied numerically.

From the results shown in Fig.~1 to Fig.~5,
we have found that the quasi-nondecoupling effect from the B-term mixing 
can be significant in the 4HDSSM, which can change the MSSM observables
$m_{H^\pm}$, $m_h$, $m_H$ and $\sin^2(\beta-\alpha)$
to a considerable extent. 
When the Higgs boson $h$ is found via the processes of
gluon fusion or vector boson fusion at the LHC,  
$m_h$ is expected to be measured very accurately.
The correction $\delta_h$ due to the quasi-nondecoupling effect can be
much larger than the expected error at the LHC and the ILC.
Therefore, we conclude that the effect on $m_h$ can be measured. 
The other MSSM Higgs bosons $H$, $A$ and $H^\pm$ are expected to be
discovered  at the LHC as long as $m_A$ is not too large. 
The deviations due to the quasi-nondecoupling effect in these quantities may also be
identified when they are precisely measured at the LHC or the ILC.

Detecting the deviations from the MSSM predictions on these MSSM
observables, the MSSM Higgs sector can be tested,
and at the same time the possibility of extended SUSY Higgs sectors
including the 4HDSSM can be explored even
when only the MSSM particles are discovered 
in near future at the LHC and at the ILC.
A detailed discussion on the Yukawa sector is given elsewhere.

\acknowledgments{
This work was supported in part by Grant-in-Aid for Scientific Research, 
No. 22740137 [M.A.], No. 22244031~[S.K.], and No. 22011007~[T.S.].
K.Y. was supported by Japan Society for the Promotion of Science.}

\appendix

\section{Rotation of the  basis}

The Higgs potential in our model is 
given from Lagrangian in Eq.~(\ref{eq:lag}) by 
\begin{align}
V_{\text{H}}^{}=&
({M}_{-}^2)_{11}^{}\hat{\Phi}_1^{\dagger}\hat{\Phi}_1^{}
+({M}_{-}^2)_{13}^{}\hat{\Phi}_1^{\dagger}\hat{\Phi}_3^{}
+({M}_{-}^2)_{13}^{*}\hat{\Phi}_3^{\dagger}\hat{\Phi}_1^{}
+({M}_{-}^2)_{33}^{}\hat{\Phi}_3^{\dagger}\hat{\Phi}_3^{}
\displaybreak[0]
\nonumber\\
&+({M}_{+}^2)_{22}^{}\hat{\Phi}_2^{\dagger}\hat{\Phi}_2^{}
+({M}_{+}^2)_{24}^{}\hat{\Phi}_2^{\dagger}\hat{\Phi}_4^{}
+({M}_{+}^2)_{24}^{\ast}\hat{\Phi}_4^{\dagger}\hat{\Phi}_2^{}
+({M}_{+}^2)_{44}^{}\hat{\Phi}_4^{\dagger}\hat{\Phi}_4^{}
\displaybreak[0]
\nonumber\\
&
+\left(
B_{12}\mu_{12} \hat{\Phi}_1^{}\cdot \hat{\Phi}_2^{}
+B_{34}\mu_{34} \hat{\Phi}_3^{}\cdot \hat{\Phi}_4^{}
+B_{14}\mu_{14} \hat{\Phi}_1^{}\cdot \hat{\Phi}_4^{}
+B_{32}\mu_{32} \hat{\Phi}_3^{}\cdot \hat{\Phi}_2^{}
+\text{h.c.}\right)
\displaybreak[0]
\nonumber\\
&
+\frac{g^{\prime 2}+g^2}{8}\left(
\hat{\Phi}_2^{\dagger}\hat{\Phi}_2^{}
+\hat{\Phi}_4^{\dagger}\hat{\Phi}_4^{}
-\hat{\Phi}_1^{\dagger}\hat{\Phi}_1^{}
-\hat{\Phi}_3^{\dagger}\hat{\Phi}_3^{}
\right)^2
\displaybreak[0]
\nonumber\\
&
+\frac{g^2}{2}\left\{
(\hat{\Phi}_1^{\dagger}\hat{\Phi}_2^{})(\hat{\Phi}_2^\dagger\hat{\Phi}_1^{})
+(\hat{\Phi}_1^{\dagger}\hat{\Phi}_4^{})(\hat{\Phi}_4^\dagger\hat{\Phi}_1^{})
+(\hat{\Phi}_3^{\dagger}\hat{\Phi}_2^{})(\hat{\Phi}_2^\dagger\hat{\Phi}_3^{})
+(\hat{\Phi}_3^{\dagger}\hat{\Phi}_4^{})(\hat{\Phi}_4^\dagger\hat{\Phi}_3^{})
\right.
\nonumber\\
&\phantom{+\frac{g^2}{2}()()}\left.
+(\hat{\Phi}_1^{\dagger}\hat{\Phi}_3^{})(\hat{\Phi}_3^{\dagger}\hat{\Phi}_1^{})
-(\hat{\Phi}_1^{\dagger}\hat{\Phi}_1^{})(\hat{\Phi}_3^{\dagger}\hat{\Phi}_3^{})
+(\hat{\Phi}_2^{\dagger}\hat{\Phi}_4^{})(\hat{\Phi}_4^{\dagger}\hat{\Phi}_2^{})
-(\hat{\Phi}_2^{\dagger}\hat{\Phi}_2^{})(\hat{\Phi}_4^{\dagger}\hat{\Phi}_4^{})
\right\}\;,
\end{align}
where
$g$ and $g'$ are the $\text{SU}(2)_L$ and $\text{U}(1)_Y$ gauge couplings 
respectively, and 
\begin{align}
(M_{-}^2)_{11}^{}=&(\tilde{M}_{-}^2)_{11}^{}+|\mu_{12}^{}|^2+|\mu_{14}^{}|^2\;,\nonumber\\
(M_{-}^2)_{33}^{}=&(\tilde{M}_{-}^2)_{33}^{}+|\mu_{32}^{}|^2+|\mu_{34}^{}|^2\;,\nonumber\\
(M_{-}^2)_{13}^{}=&(\tilde{M}_{-}^2)_{13}^{}+\mu_{12}^{*}\mu_{32}^{}+\mu_{14}^{*}\mu_{34}^{}\;,\nonumber\\
(M_{+}^2)_{22}^{}=&(\tilde{M}_{+}^2)_{22}^{}+|\mu_{12}^{}|^2+|\mu_{32}^{}|^2\;,\nonumber\\
(M_{+}^2)_{44}^{}=&(\tilde{M}_{+}^2)_{44}^{}+|\mu_{14}^{}|^2+|\mu_{34}^{}|^2\;,\nonumber\\
(M_{+}^2)_{24}^{}=&(\tilde{M}_{-}^2)_{24}^{}+\mu_{12}^{*}\mu_{14}^{}+\mu_{32}^{*}\mu_{34}^{}\;.
\end{align}

%\subsection{The gauge eigenstate basis}

Because $\hat{\Phi}_1^{}$ and $\hat{\Phi}_3^{}$, and $\hat{\Phi}_2^{}$
and $\hat{\Phi}_4^{}$ 
have respectively the same quantum numbers, 
we may rotate the basis by introducing 
%the mixing angles $\theta_+$ and $\theta_-$ as 
the $2\times 2$ unitary mixing matrices $U_+$ and $U_-$ as 
\begin{equation}
\begin{pmatrix}
\hat{\Phi}_1^{}\\
\hat{\Phi}_3^{}
\end{pmatrix}
\to 
\begin{pmatrix}
{\Phi}_1^{}\\
{\Phi}_1^{\prime}
\end{pmatrix}
=
U_-
\begin{pmatrix}
\hat{\Phi}_1^{}\\
\hat{\Phi}_3^{}
\end{pmatrix}\;,\quad
\begin{pmatrix}
\hat{\Phi}_2^{}\\
\hat{\Phi}_4^{}
\end{pmatrix}
\to 
\begin{pmatrix}
\Phi_2^{}\\
\Phi_2^{\prime}
\end{pmatrix}
=
U_+
\begin{pmatrix}
\hat{\Phi}_2^{}\\
\hat{\Phi}_4^{}
\end{pmatrix}\;.
\end{equation}
By using this degrees of freedom, one may choose the basis where
only ${\Phi}_1^{}$ and ${\Phi}_2^{}$ have VEV's while  
those of ${\Phi}_1^{\prime}$ and ${\Phi}_2^{\prime}$ are zero.
In this basis, the Higgs potential is expressed by 
\begin{align}
V_{\text{H}}^{}=&
\begin{pmatrix}
{\Phi}_1^{\dagger}&{\Phi}_1^{\prime\dagger}
\end{pmatrix}
U_-
\begin{pmatrix}
(M_{-}^{2})_{11}^{}&
(M_{-}^{2})_{13}^{}\\
(M_{-}^{2})_{13}^{*}&
(M_{-}^{2})_{33}^{}\\
\end{pmatrix}
U_-^{\dagger}
\begin{pmatrix}
{\Phi}_1^{} \\ {\Phi}_1^{\prime}
\end{pmatrix}
+
\begin{pmatrix}
{\Phi}_2^{\dagger}&{\Phi}_2^{\prime\dagger}
\end{pmatrix}
U_+
\begin{pmatrix}
(M_{+}^{2})_{22}^{}&
(M_{+}^{2})_{24}^{}\\
(M_{+}^{2})_{24}^{*}&
(M_{+}^{2})_{44}^{}\\
\end{pmatrix}
U_+^{\dagger}
\begin{pmatrix}
{\Phi}_2^{} \\ {\Phi}_2^{\prime}
\end{pmatrix}
\displaybreak[0]
\nonumber\\
&
+\left(
\begin{pmatrix}
{\Phi}_1^{}&
{\Phi}_1^{\prime}
\end{pmatrix}
U_-^*
\begin{pmatrix}
B_{12}^{}\mu_{12}^{}&
B_{14}^{}\mu_{14}^{}\\
B_{32}^{}\mu_{32}^{}&
B_{34}^{}\mu_{34}^{}\\
\end{pmatrix}
U_+^{\dagger}
\cdot 
\begin{pmatrix}
{\Phi}_2^{}\\
{\Phi}_2^{\prime}
\end{pmatrix}
+\text{h.c.}\right)
\displaybreak[0]
\nonumber\\
&
+\frac{g^{\prime 2}+g^2}{8}\left(
{\Phi}_2^{\dagger}{\Phi}_2^{}
+{\Phi}_2^{\prime\dagger}{\Phi}_2^{\prime}
-{\Phi}_1^{\dagger}{\Phi}_1^{}
-{\Phi}_1^{\prime\dagger}{\Phi}_1^{\prime}
\right)^2
\displaybreak[0]
\nonumber\\
&
+\frac{g^2}{2}\left\{
({\Phi}_1^{\dagger}{\Phi}_2^{})({\Phi}_2^\dagger{\Phi}_1^{})
+({\Phi}_1^{\dagger}{\Phi}_2^{\prime})({\Phi}_2^{\prime\dagger}{\Phi}_1^{})
+({\Phi}_1^{\prime\dagger}{\Phi}_2^{})({\Phi}_2^\dagger{\Phi}_1^{\prime})
+({\Phi}_1^{\prime\dagger}{\Phi}_2^{\prime})({\Phi}_2^{\prime\dagger}{\Phi}_1^{\prime})
\right.
\nonumber\\
&\phantom{+\frac{g^2}{2}()()}\left.
+(\Phi_1^{\dagger}\Phi_1^{\prime})(\Phi_1^{\prime\dagger}\Phi_1^{})
-(\Phi_1^{\dagger}\Phi_1^{})(\Phi_1^{\prime\dagger}\Phi_1^{\prime})
+(\Phi_2^{\dagger}\Phi_2^{\prime})(\Phi_2^{\prime\dagger}\Phi_2^{})
-(\Phi_2^{\dagger}\Phi_2^{})(\Phi_2^{\prime\dagger}\Phi_2^{\prime})
\right\}\;.
\end{align}
Hereafter, we reparameterize the parameters as 
\begin{align}
&
U_-^{}
\begin{pmatrix}
(M_{-}^{2})_{11}^{}&
(M_{-}^{2})_{13}^{}\\
(M_{-}^{2})_{13}^{*}&
(M_{-}^{2})_{33}^{}\\
\end{pmatrix}
U_-^{\dagger}
\to 
\begin{pmatrix}
(M_{1}^{2})_{11}^{}&
(M_{1}^{2})_{12}^{}\\
(M_{1}^{2})_{12}^{*}&
(M_{1}^{2})_{22}^{}\\
\end{pmatrix}\;,\nonumber\\
&
U_+^{}
\begin{pmatrix}
(M_{+}^{2})_{22}^{}&
(M_{+}^{2})_{24}^{}\\
(M_{+}^{2})_{24}^{*}&
(M_{+}^{2})_{44}^{}\\
\end{pmatrix}
U_+^{\dagger}
\to 
\begin{pmatrix}
(M_{2}^{2})_{11}^{}&
(M_{2}^{2})_{12}^{}\\
(M_{2}^{2})_{12}^{*}&
(M_{2}^{2})_{22}^{}\\
\end{pmatrix}\;,\nonumber\\
&
-U_-^*
\begin{pmatrix}
B_{12}^{}\mu_{12}^{}&
B_{14}^{}\mu_{14}^{}\\
B_{32}^{}\mu_{32}^{}&
B_{34}^{}\mu_{34}^{}\\
\end{pmatrix}
U_+^{\dagger}
\to
\begin{pmatrix}
(M_{3}^{2})_{11}^{}&
(M_{3}^{2})_{12}^{}\\
(M_{3}^{2})_{21}^{}&
(M_{3}^{2})_{22}^{}\\
\end{pmatrix}\;.
\end{align}
With the above notation, the Higgs potential can be rewritten
as in Eq.~(\ref{V_gi}).

\section{Yukawa Interactions}

In the gauge eigenstate basis, the Yukawa coupling matrices
$(Y_f)_{ij}$ and $(Y_f^{\prime})_{ij}$
$(f=u,d,e)$ associated with
$\Phi_k^{}$ and $\Phi_k^{\prime}$ ($k=1$ for $f=d,e$ and $k=2$ for $f=u$) are
given as
\begin{align}
(Y_u)_{ij}=&
(U_+)_{11}^*(\hat{Y}_u)_{ij}+(U_+)_{12}^*(\hat{Y}_u^{\prime})_{ij}\;,\quad
(Y_d)_{ij}=
(U_-)_{11}^*(\hat{Y}_d)_{ij}+(U_-)_{12}^*(\hat{Y}_d^{\prime})_{ij}\;,\quad\nonumber\\
(Y_e)_{ij}=&
(U_-)_{11}^*(\hat{Y}_e)_{ij}+(U_-)_{12}^*(\hat{Y}_e^{\prime})_{ij}\;,\quad\nonumber\\
(Y_u^{\prime})_{ij}=&
(U_+)_{21}^*(\hat{Y}_u)_{ij}+(U_+)_{22}^*(\hat{Y}_u^{\prime})_{ij}\;,\quad
(Y_d^{\prime})_{ij}=
(U_-)_{21}^*(\hat{Y}_d)_{ij}+(U_-)_{22}^*(\hat{Y}_d^{\prime})_{ij}\;,\quad\nonumber\\
(Y_e^{\prime})_{ij}=&
(U_-)_{21}^*(\hat{Y}_e)_{ij}+(U_-)_{22}^*(\hat{Y}_e^{\prime})_{ij}\;.
\end{align}
The formulae here are given in the basis where $Y_u$ and $Y_d$ are diagonal, {\it i.e.},
\begin{equation}
(Y_u)_{ij}=\frac{\sqrt{2}m_{u_i}}{v\sin\beta}\delta_{ij}\;,\quad
(Y_d)_{ij}=\frac{\sqrt{2}m_{d_i}}{v\cos\beta}V_{ji}^*\;,\quad
(Y_e)_{ij}=\frac{\sqrt{2}m_{e_i}}{v\cos\beta}\delta_{ij}\;,
\end{equation}
where 
$m_{u_i}$, $m_{d_i}$ and $m_{e_i}$ are the masses of up-type quarks, down-type quarks, and the charged leptons, respectively, 
and $V_{ij}$ is the Cabbibo-Kobayashi-Maskawa matrix.
The fields $f_{Ri}$ and $f_{Li}$ $(f=u,d,e)$ denote the mass eigenstates, and the left-handed component of the 
quark mass eigenstates are embedded into the SU(2) doublets as
\begin{equation}
q_{Li}^{}=
\begin{pmatrix}
u_{Li}^{}\\
V_{ij}d_{Lj}\;.
\end{pmatrix}
\end{equation}
From the superpotential given in Eq.~(\ref{eq:superpotential}), 
the Yukawa interactions between the matter fields and the neutral Higgs scalar fields are 
read as follows:
\begin{align}
\mathcal{L}=& 
-\left(\frac{m_{u_i}\delta_{ij}}{vs_{\beta}}(O_H)_{2\alpha }+\frac{(Y_u^{\prime})_{ij}}{\sqrt{2}}(O_H)_{4\alpha}\right)\bar{u}_{Ri}u_{Lj}\varphi^{\text{even}}_{\alpha}
\nonumber\\
&-\left(\frac{m_{d_i}\delta_{ij}}{vc_{\beta}}(O_H)_{1\alpha}+\frac{(\tilde{Y}_d^{\prime})_{ij}}{\sqrt{2}}(O_H)_{3\alpha}\right)\bar{d}_{Ri}d_{Lj}\varphi^{\text{even}}_{\alpha}
\nonumber\\
&
-\left(\frac{m_{e_i}\delta_{ij}}{vc_{\beta}}(O_H)_{1\alpha}+\frac{(Y_{e}^{\prime})_{ij}}{\sqrt{2}}(O_H)_{3\alpha}\right)\bar{e}_{Ri}e_{Lj}\varphi^{\text{even}}_{\alpha}
\nonumber\\
&
-i\left(\frac{m_{u_i}\delta_{ij}}{v}\left(\cot\beta (\bar{O}_A)_{2\alpha}-(\bar{O}_A)_{1\alpha}\right)
+\frac{(Y_u^{\prime})_{ij}}{\sqrt{2}}(\bar{O}_A)_{4\alpha}\right)\bar{u}_{Ri}u_{Lj}\varphi^{\text{odd}}_{\alpha}
\nonumber\\
&
-i\left(\frac{m_{d_i}\delta_{ij}}{v}
\left(\tan\beta (\bar{O}_A)_{2\alpha}+(\bar{O}_A)_{1\alpha}\right)
+\frac{(\tilde{Y}_d^{\prime})_{ij}}{\sqrt{2}}(\bar{O}_H)_{3\alpha}\right)\bar{d}_{Ri}d_{Lj}\varphi^{\text{odd}}_{\alpha}
\nonumber\\
&
-i\left(\frac{m_{e_i}\delta_{ij}}{v}
\left(\tan\beta (\bar{O}_A)_{2\alpha}+(\bar{O}_A)_{1\alpha}\right)
+\frac{(Y_{e}^{\prime})_{ij}}{\sqrt{2}}(\bar{O}_A)_{3\alpha}\right)\bar{e}_{Ri}e_{Lj}\varphi^{\text{odd}}_{\alpha}
+\text{h.c.}
\;,
\end{align}
where we parameterize the extra down-type Yukawa couplings as
$(Y_d^{\prime})_{ij}=(\tilde{Y}_d^{\prime})_{ik}V_{jk}$,  
CP-even and CP-odd Higgs scalar bosons are written as 
$\varphi^{\text{even}}_{\alpha}=(H, h, H_1^{\prime}, H_2^{\prime})$ and 
$\varphi^{\text{odd}}_{\alpha}=(z^0, A, A_1, A_2)$ with $z^0$ denoting a Nambu-Goldstone mode, 
and the mixing matrix $O_H$ and $\bar{O}_A$ are defined as 
\begin{equation}
O_H^{T}M_H^2O_H^{}=
\begin{pmatrix}
m_H^2&0&0&0\\
0&m_h^2&0&0\\
0&0&m_{H_1'}^2&0\\
0&0&0&m_{H_2'}^2
\end{pmatrix}\;,\quad
\bar{O}_A^{T}\bar{M}_A^2\bar{O}_A^{}=
\begin{pmatrix}
0&0&0&0\\
0&m_A^2&0&0\\
0&0&m_{A_1}^2&0\\
0&0&0&m_{A_2}^2
\end{pmatrix}\;,
\end{equation}
with 
$(\bar{O}_A)_{1\alpha}=(\bar{O}_A)_{\alpha 1}=\delta_{1\alpha}$.

For the Yukawa interactions with a charged scalar fields, they are found as
\begin{align}
\mathcal{L}=&
\left(\frac{\sqrt{2}m_{u_i}\delta_{ik}}{v}\left(\cot\beta(\bar{O}_{\pm})_{2\alpha}-(\bar{O}_{\pm})_{1\alpha}\right)
+(Y_u^{\prime})_{ik}(\bar{O}_{\pm})_{4\alpha}\right)V_{kj}\bar{u}_{Ri}d_{Lj}\varphi^+_{\alpha}
\nonumber\\
&
+\left(\frac{\sqrt{2}m_{d_k}\delta_{kj}}{v}\left(\tan\beta(\bar{O}_{\pm})_{2\alpha}+(\bar{O}_{\pm})_{1\alpha}\right)
+(\tilde{Y}_d^{\prime *})_{jk}(\bar{O}_{\pm})_{3\alpha}\right)V_{ik}\bar{u}_{Li}d_{Rj}\varphi^+_{\alpha}
\nonumber\\
&
+\left(\frac{\sqrt{2}m_{e_j}\delta_{ij}}{v}\left(\tan\beta(\bar{O}_{\pm})_{2\alpha}+(\bar{O}_{\pm})_{1\alpha}\right)
+(Y_e^{\prime *})_{ji}(\bar{O}_{\pm})_{3\alpha}\right)\bar{\nu}_{Li}d_{Rj}\varphi^+_{\alpha}
+\text{h.c.}
\;,
\end{align}
where 
the charged scalar bosons are written as $\varphi^+_{\alpha}=(w^+,H^+,H_1^+,H_2^+)$ with $w^+$ being 
a Nambu-Goldstone boson, and the mixing matrix $\bar{O}_{\pm}$ is defined as 
\begin{equation}
\bar{O}_{\pm}^T\bar{M}_{H^{\pm}}^2\bar{O}_{\pm}
=\begin{pmatrix}
0&0&0&0\\
0&m_{H^{\pm}}^2&0&0\\
0&0&m_{H_1^{\pm}}^2&0\\
0&0&0&m_{H_2^{\pm}}^2\\
\end{pmatrix}\;,
\end{equation}
with $(\bar{O}_\pm)_{1\alpha}=(\bar{O}_\pm)_{\alpha 1}=\delta_{1\alpha}$.

In the decoupling region $m_A^{}/M \ll 1$, the Yukawa interactions with the MSSM Higgs bosons are given as 
\begin{align}
\mathcal{L}=
&-\left\{
	\frac{m_{u_i}\delta_{ij}}{v}\left(\sin(\beta-\alpha_{\text{eff}})+R\cot\beta\cos(\beta-\alpha_{\text{eff}})\right)
	\right.\nonumber\\
	&\left.~~~~~~-\frac{(Y_u^{\prime})_{ij}^{}}{\sqrt{2}}\frac{|k|R}{r}\cos(\beta-\alpha_{\text{eff}})
	+\mathcal{O}\left(\frac{m_A^2}{M^2}\right)
	\right\} \bar{u}_{Ri}u_{Lj}h
	\displaybreak[0]
	\nonumber\\
&-\left\{
	\frac{m_{d_i}\delta_{ij}}{v}\left(\sin(\beta-\alpha_{\text{eff}})-R\tan\beta\cos(\beta-\alpha_{\text{eff}})\right)
	\right.\nonumber\\
	&\left.~~~~~~-\frac{(\tilde{Y}_d^{\prime})_{ij}^{}}{\sqrt{2}}
	|k'| R \cos(\beta-\alpha_{\text{eff}})
	+\mathcal{O}\left(\frac{m_A^2}{M^2}\right)
	\right\}\bar{d}_{Ri}d_{Lj}h
	\displaybreak[0]
	\nonumber\\
&-\left\{
	\frac{m_{e_i}\delta_{ij}}{v}\left(\sin(\beta-\alpha_{\text{eff}})-R\tan\beta\cos(\beta-\alpha_{\text{eff}})\right)
	\right.\nonumber\\
	&\left.~~~~~~-\frac{(Y_e^{\prime})_{ij}^{}}{\sqrt{2}}
	|k'| R \cos(\beta-\alpha_{\text{eff}})
	+\mathcal{O}\left(\frac{m_A^2}{M^2}\right)
	\right\}\bar{e}_{Ri}e_{Lj}h
	\displaybreak[0]
	\nonumber\\
&-\left\{
	\frac{m_{u_i}\delta_{ij}}{v}\left(\cos(\beta-\alpha_{\text{eff}})-R\cot\beta\sin(\beta-\alpha_{\text{eff}})\right)
	\right.\nonumber\\
	&\left.~~~~~~+\frac{(Y_u^{\prime})_{ij}^{}}{\sqrt{2}}\frac{|k|R}{r}\sin(\beta-\alpha_{\text{eff}})
	+\mathcal{O}\left(\frac{m_A^2}{M^2}\right)
	\right\}\bar{u}_{Ri}u_{Lj}H
	\displaybreak[0]
	\nonumber\\
&-\left\{
	\frac{m_{d_i}\delta_{ij}}{v}\left(\cos(\beta-\alpha_{\text{eff}})+R\tan\beta\sin(\beta-\alpha_{\text{eff}})\right)
	\right.\nonumber\\
	&\left.~~~~~~+\frac{(\tilde{Y}_d^{\prime})_{ij}^{}}{\sqrt{2}}|k'| R \sin(\beta-\alpha_{\text{eff}})
	+\mathcal{O}\left(\frac{m_A^2}{M^2}\right)
	\right\}\bar{d}_{Ri}d_{Lj}H
	\displaybreak[0]
	\nonumber\\
&-\left\{
	\frac{m_{e_i}\delta_{ij}}{v}\left(\cos(\beta-\alpha_{\text{eff}})+R\tan\beta\sin(\beta-\alpha_{\text{eff}})\right)
	\right.\nonumber\\
	&\left.~~~~~~+\frac{(Y_e^{\prime})_{ij}^{}}{\sqrt{2}}|k'| R \sin(\beta-\alpha_{\text{eff}})
	+\mathcal{O}\left(\frac{m_A^2}{M^2}\right)
	\right\}\bar{e}_{Ri}e_{Lj}H
	\displaybreak[0]
	\nonumber\\
&-i\left\{
	\frac{m_{u_i}\delta_{ij}}{v}R\cot\beta 
	-\frac{(Y_u^{\prime})_{ij}^{}}{\sqrt{2}}\frac{|k|R}{r}
	+\mathcal{O}\left(\frac{m_A^2}{M^2}\right)
	\right\}\bar{u}_{Ri}u_{Lj}A
	\displaybreak[0]
	\nonumber\\
&-i\left\{
	\frac{m_{d_i}\delta_{ij}}{v}  R\tan\beta
	+\frac{(\tilde{Y}_d^{\prime})_{ij}^{}}{\sqrt{2}}|k'| R
	+\mathcal{O}\left(\frac{m_A^2}{M^2}\right)
	\right\}\bar{d}_{Ri}d_{Lj}A
	\displaybreak[0]
	\nonumber\\
&-i\left\{
	\frac{m_{e_i}\delta_{ij}}{v}R \tan\beta
	+\frac{({Y}_e^{\prime})_{ij}^{}}{\sqrt{2}}|k'| R 
	+\mathcal{O}\left(\frac{m_A^2}{M^2}\right)
	\right\}\bar{e}_{Ri}e_{Lj}A
	\displaybreak[0]
	\nonumber\\
&+\left\{
	\frac{\sqrt{2}m_{u_i}\delta_{ik}}{v}R\cot\beta -(Y_u^{\prime})_{ik}\frac{|k|R}{r}
	\right\}V_{kj}\bar{u}_{Ri}d_{Lj}H^+
	\displaybreak[0]
	\nonumber\\
&+\left\{
	\frac{\sqrt{2}m_{d_i}\delta_{ik}}{v}R\tan\beta
 +(\tilde{Y}_d^{\prime})_{ik} |k'| R 
	\right\}V_{kj}\bar{u}_{Li}d_{Rj}H^+
	\displaybreak[0]
	\nonumber\\
&+\left\{
	\frac{\sqrt{2}m_{e_i}\delta_{ij}}{v}R\tan\beta +(Y_e^{\prime})_{ij}
 |k'| R
	\right\}\bar{\nu}_{Li}e_{Rj}H^+
	+\text{h.c.}\;.
\end{align}
In the SM limit where both $M$ and $m_A$ are enough heavy compared to $m_Z$, the contributions from $Y_f^{\prime}$ disappear.
On the other hand, effects of extra Yukawa contribution can remain in the Yukawa couplings with $H$, $A$, and $H^{\pm}$
as quasi-nondecoupling effects, even if $M$ is much larger than $m_A$.
In general, $Y_f^{\prime}$ has non-trivial flavor structure and the flavor changing processes are enhanced a lot, if 
there are off-diagonal elements in $(Y_u)_{ij}$, $(\tilde{Y}_d)_{ij}$ and $(Y_e)_{ij}$.
In order to suppress a dangerous contributions to the flavor changing processes, 
some mechanism to forbid such the off-diagonal elements is necessary unless $m_A$ is very large.
Discrete symmetries such as $Z_2$ symmetry are often considered.
The way of the $Z_2$ parity assignment are discussed in Sec.~II, and
the possible types are listed in Table~1.
It is interesting to study how the extra Yukawa interaction contribute to
flavor measurements in these types of Yukawa interaction in the 4HDSSM.
Further discussions on flavor physics are given elsewhere~\cite{aksy-flavor}.

%\vspace*{-4mm}


\begin{thebibliography}{1}
\bibitem{comment}
We note that there is also a possibility that extra doublet fields do not 
have VEVs by imposing an exact (unbroken) discrete $Z_2$ symmetry~\cite{susy-ma}.
\end{thebibliography}
\end{document}